\documentclass[fleqn,usenatbib]{mnras}

\usepackage{newtxtext,newtxmath}

\usepackage[T1]{fontenc}
\usepackage{adjustbox}
\usepackage{blindtext}
\DeclareRobustCommand{\VAN}[3]{#2}
\let\VANthebibliography\thebibliography
\def\thebibliography{\DeclareRobustCommand{\VAN}[3]{##3}\VANthebibliography}

\usepackage{comment}
\usepackage{graphicx}	
\usepackage{amsmath}	
\usepackage{dcolumn}
\usepackage{hyperref}
\usepackage{kantlipsum}
\usepackage{dblfloatfix}
\usepackage{longtable}
\usepackage{array}
\usepackage{caption}
\usepackage{threeparttable}
\usepackage{url}
\usepackage{subcaption}
\newcolumntype{M}[1]{>{\centering\arraybackslash}m{#1}}
\newcolumntype{N}{@{}m{0pt}@{}}

\usepackage{bm}

\title [AstroSat observations of GRS 1915+105] {Correlations between QPO frequencies and  spectral parameters of GRS 1915+105 using AstroSat observations}

\author[Dhaka et al.]{Ruchika Dhaka,$^{1}$\thanks{E-mail: ruchika@iitk.ac.in, ruchikadhaka1997@gmail.com}
Ranjeev Misra,$^{2}$\thanks{rmisra@iucaa.ac.in}
J.S. Yadav$^{1,3}$\thanks{jsyadavtifr@gmail.com}
and Pankaj Jain$^{1}$\thanks{pkjain@iitk.ac.in}
\\
$^{1}$Department of physics, IIT Kanpur, Kanpur, Uttar Pradesh
208016, India\\
$^{2}$Inter-University Center for Astronomy and Astrophysics,
Ganeshkhind, Pune 411007, India\\
$^{3}$Tata Institute of Fundamental Research, Homi Bhabha Road, Mumbai, India 
}

\date{Accepted XXX. Received YYY; in original form ZZZ}

\pubyear{2022}

\begin{document}
\label{firstpage}
\pagerange{\pageref{firstpage}--\pageref{lastpage}}
\maketitle

\begin{abstract}
In this work, we study the correlation between Quasi-periodic Oscillation (QPO) frequency and the spectral parameters during various X-ray states in the black hole binary GRS 1915+105 which matches well with the predicted relativistic dynamic frequency (i.e. the inverse of the sound crossing time) at the truncated radii.  We have used broadband data of LAXPC and SXT instruments onboard AstroSat. Spectral fitting shows that the accretion rate varies from $\sim 0.1$ to $\sim 5.0 \times 10^{18}$ gm/s and the truncated radius changing from the last stable orbit of an almost maximally spinning black hole, $\sim$ 1.2  to $\sim$ 19 Gravitational radii. For this wide range, the frequencies of the C-type QPO (2 - 6 Hz) follow the trend predicted by the relativistic dynamical frequency model and interestingly, the high-frequency QPO at $\sim$ 70 Hz also follows the same trend, suggesting they originate from the innermost stable circular orbit with the same mechanism as the more commonly observed C-type QPO. While the qualitative trend is as predicted, there are quantitative deviations between the data and the theory, and the possible reasons for these deviations are discussed.
\end{abstract}

\begin{keywords}
accretion, accretion discs - black hole physics - stars: black holes - X-rays: binaries - relativistic processes
\end{keywords}



\section{Introduction}
\label{sec:intro}
The Black Hole X-ray Binary (BHXB) GRS 1915+105 was discovered on August 15, 1992, as a transient by the WATCH All-sky monitor onboard Granat observatory. It was the first galactic object to show a superluminal jet \citep[][]{mirabel1994superluminal}. The binary system contains a black hole of 12.4 solar mass \citep[][]{reid2014parallax}. This source is located at a distance D = 8.6 kpc \citep{reid2014parallax} and its relativistic jets are directed at an angle $i=70^\circ$ from the line of sight \citep{mirabel1994superluminal}.  It is an outstanding source because of its huge variability \citep{castro1992grs, belloni2000model,belloni1997unified, yadav1999different}. This source is observed in 14 different X-ray classes, based on its X-ray flux, Color-Color Diagram (CCD) and hardness ratio \citep[][]{belloni2000model,klein2002hard, hannikainen2005characterizing}. Some of these classes are named  $\phi$, $\chi$, $\theta$, $\lambda$, $\rho$, etc. Among all the 14 different classes the most observed class is $\chi$. The $\chi$ class is the least variable class, and no large amplitude and long-term X-ray flux variability have been observed. Most of the time, since its discovery in 1992, GRS 1915+105 has been seen in bright X-ray states like High Soft state (HS) and High HIMS state (also called the Steep Power Law state (SPL state)). This source has entered into a decline phase since 2018 (lower branch of HIMS and the Low Hard State (LS)).

\begin{table*}
\centering
\caption{Details of observations of the source GRS 1915+105 made by AstroSat between 2016 and 2019. In the table, observation IDs are listed alongside the exposure time, date, and time of the observation.}
\label{tab:table1}
\begin{tabular}{c c c c c c} 
\hline \hline
Observation No. & Observation Date  & Observation ID & Start Time(hh:mm:ss) & LAXPC Exposure Time(ks) & SXT Exposure Time(ks) \\
 
\hline
1 & 03 Mar 2016 (MJD 57450) & T01\_030T01\_9000000358  & 09:54:20 & 22.840  & 7.822 \\  
2 & 25 Apr 2016 (MJD 57503) &  G05\_214T01\_9000000428  & 03:57:56 & 8.209   & 3.045 \\ 
3 & 27 Apr 2016 (MJD 57505) & G05\_167T02\_9000000432 & 17:22:26 & 14.440  & 7.358 \\
4 & 28 Mar 2017 (MJD 57840) & G06\_033T01\_9000001116  & 18:03:19 & 32.050   & 13.190\\
5 & 01 Apr 2017 (MJD 57844) & G07\_046T01\_9000001124  & 11:50:10 & 11.420  &  5.452\\
 
6  & 15 Apr 2017 (MJD 57858) & G07\_028T01\_9000001166 & 22:39:28 & 9.668  & 4.974\\
7 & 21 Mar 2019 (MJD 58563) & A05\_173T01\_9000002812 & 19:14:36 & 63.498 & 20.929\\
 \hline
\end{tabular}
\end{table*}

\begin{figure*}
    \centering
\includegraphics[scale=0.5]{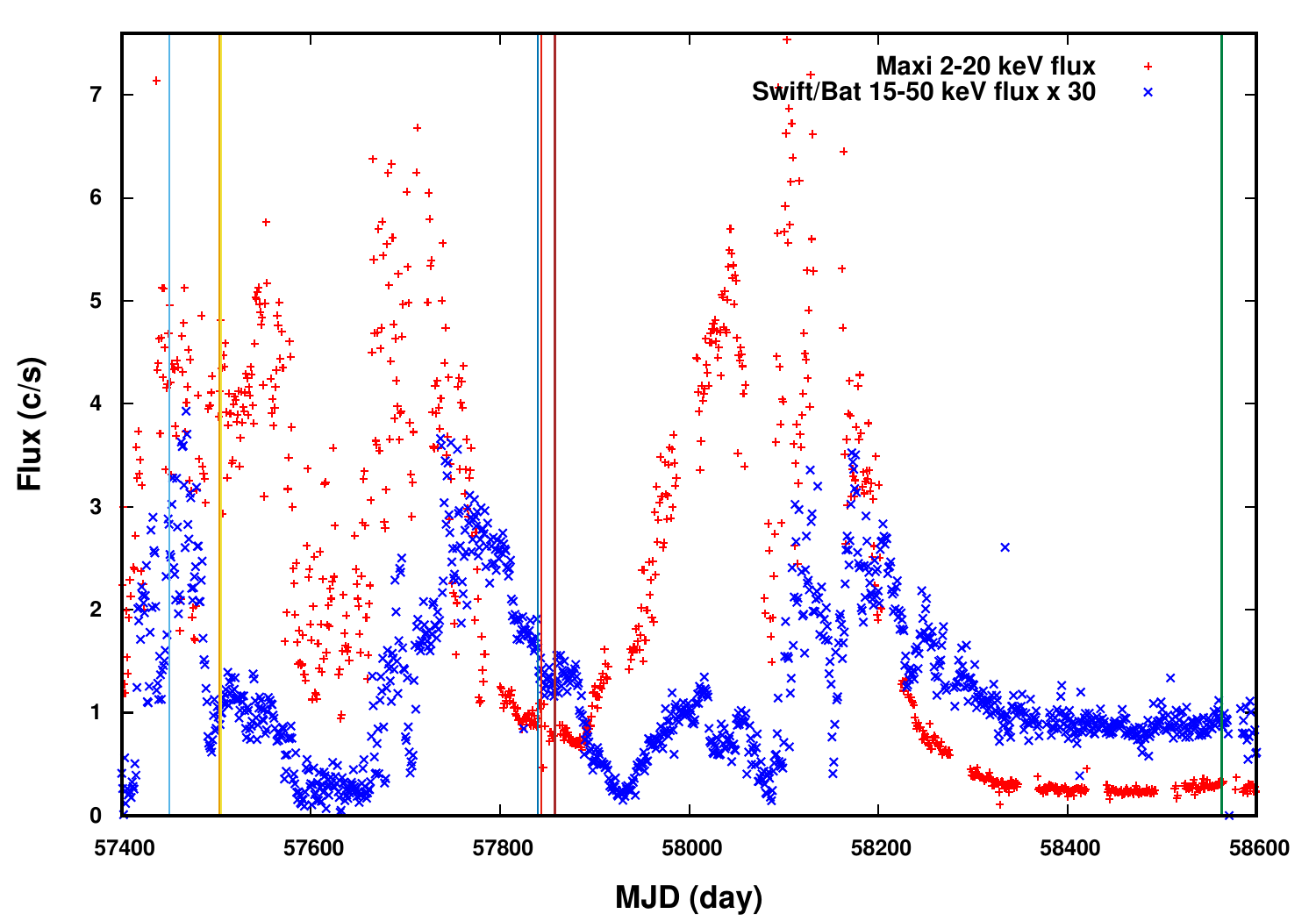}
    \caption{1 day binned complete lightcurve of GRS 1915+105 using MAXI and Swift/BAT data starting from 13  January 2016 (57400 MJD) to 27  April 2019 (58600 MJD). The vertical solid lines represent AstroSat observations taken for the analysis, with dates provided in Table \ref{tab:table1}. (The vertical lines for the 25 \& 27 of April 2016, at around 57500 MJD, nearly overlap and appear to be one line.)}
    
    \label{fig:fig1}
\end{figure*}

X-ray binaries exhibit variability on rapid time scales. Fourier analysis is often used to study fast variability and quasi-periodic oscillations (QPOs) by computing power-density spectra (PDS)\citep[][]{klis1989fourier}.
There are numerous patterns have been observed in the PDS \citep[][]{belloni2002unified, belloni1997energy}, ranging from various types of broad-band noise to much narrower structures known as QPOs. These appear as sharp peaks in the power spectrum. QPOs with frequencies ranging from few mHz to $\sim70$ Hz have been observed for the source GRS 1915+105 \citep[][]{morgan1997rxte, belloni2013high, paul1997quasi, yadav1999different, sreehari2020astrosat}. 
The centroid frequencies of these QPOs during specific spectral states and transitions can be associated with physical processes occurring in these systems.
Typically, there are two types of QPOs. Low-frequency QPOs have a centroid frequency $\lesssim$ 30 Hz, whereas high-frequency QPOs have a centroid frequency $\gtrsim$ 60 Hz (up to a few hundred hertz) \citep[][]{belloni2009states, belloni2013discovery}. Low-frequency QPOs are further subdivided into A, B, and C-type QPOs based on differences in power spectral properties and phase lag behavior, and they occur in various spectral states \citep[][]{homan2001correlated, remillard2002characterizing, wijnands1999complex, casella2004study}.
However, the precise physical origin of QPOs in BHXBs is so far not well understood.

\citet{misra2020identification} have studied the dependence of QPO frequency $f$ on the inner radius $r$ of the truncated accretion disk. They found that $f/\dot{M}$ is well correlated with $r$, where $\dot M$ is the accretion rate. Remarkably, the relationship between the two is well described in terms of dynamical frequency arising due to normal modes of disk oscillations \citep{misra2020identification}. 
The dynamical frequency is defined as the inverse of the sound crossing time ($f_{dyn}  \sim c_s(r)/r$). The sound crossing time is the ratio of the truncation radius and the sound speed at the inner disc. According to the standard relativistic disc model proposed by \citet{novikov1973astrophysics}, the sound speed is dependent on several factors, including the mass accretion rate ($\dot{M}$), spin, and inner radius ($r$) of the disc. This leads to the following formula for the dynamical frequency \citep[][]{misra2020identification}:

\begin{equation}
  \frac{f_{dyn}}{\dot{M}} = N \, 8979 \, \text{Hz} \, \left(\frac{r}{r_g}\right)^{-2.5} \left(\frac{M}{12.4 M_{\odot}}\right)^{-2} \times \, A^1 B^{-2} D^{-0.5} E^{-0.5} L
  \label{primeqn}
\end{equation}
where $ r_g = GM/c^2 $ is the gravitational radius, and r is the inner disc radii, $N$ is a normalisation factor to take into account the assumptions made in the standard accretion disc theory. 
\begin{figure*}
    \centering
\includegraphics[scale=0.5]{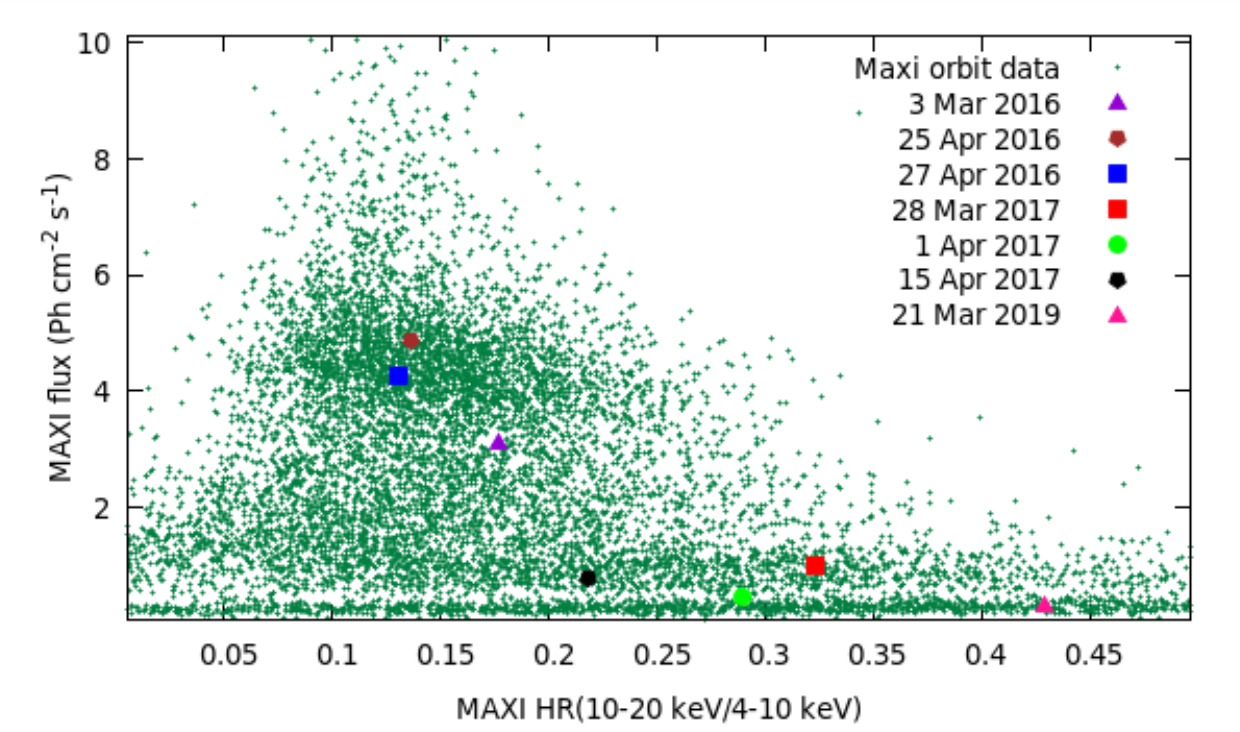}
    \caption{Hardness–intensity diagram (HID) of GRS 1915+105 showing the evolution of hardness ratio (10–20 keV/ 4-10 keV) with MAXI 2–20 keV flux from 13 Jan 2016 (MJD 57400) to 27 Apr 2019 (MJD 58600). The diagram highlights the AstroSat observations we used.}
    
    \label{fig:fig2}
\end{figure*}
The parameters $A$, $B$, $D$, $E$, and $L$ are functions of the inner disc radii and the spin parameter described in \citet{novikov1973astrophysics} and \citet{page1974disk}. All these parameters are important for small radii, $r < 10 \, r_{g}$. As a result, in this regime, the functional form of $f_{dyn}$ considerably differs from its Newtonian dependence. Using spectral and timing analysis, one can determine the mass accretion rate, inner disc radii, and QPO frequency. Thus, the interpretation, and in particular Eqn \ref{primeqn} can be verified with such an analysis. \citet{misra2020identification} did such an analysis using AstroSat observation data collected on 28 March 2016 and 1 April 2017 when GRS 1915+105 was in the low HIMS state (i.e., the lower horizontal track of HIMS). The source showed C-type QPOs in the frequency range of 3.5–5.4 Hz during the observation. A similar analysis was undertaken for Insight-HXMT observations of GRS 1915+105 when it exhibited low-frequency C-type QPOs \citep[][]{liu2021testing}. For a wider range of QPO frequency,  2.6-4.3 Hz, and  inferred accretion rate of  0.2-1.2$\times 10^{18} \textrm{gm/s}$, they confirmed the results obtained by \citet{misra2020identification}.

Apart from these C-type QPOs, GRS 1915+105 also shows a QPO at $\sim 69$ Hz, which is remarkable in having a nearly constant frequency \citep[][]{morgan1997rxte,belloni2001high,belloni2006high,belloni2013high}. This QPO has also been reported for AstroSat data, where it varied slightly from 67.4 to 72.3 Hz \citep{belloni2019variable}.

In this paper, we perform an extensive spectro-temporal analysis of various X-ray states observed in  GRS 1915+105 using AstroSat data. In GRS~1915+105, so far, only one outburst (started in 1992) is observed which is still continuing.  GRS~1915+105 is never seen in the rising phase of  an outburst. Our data includes a low hard state (Obs. 7), which has never been reported before. The motivation here is to study the QPO frequency dependence on spectral parameters covering a wider range of inner disc radii, accretion rates and QPO frequencies.

In Section \ref{sec:obs_data_red} of this work, we describe observations and data reduction techniques using the LAXPC and SXT pipeline software. In Section \ref{sec:data_analysis}, we explain the various analytical and modelling techniques used to analyse the temporal and spectral features of GRS 1915+105. In Section \ref{sec:results} of the paper, we describe the outcomes of the study and draw conclusions based on those results.


\section{Observation and Data Reduction}
\label{sec:obs_data_red}

AstroSat is a multi-wavelength observatory launched for astronomical studies of various celestial objects in near and far UV, soft (0.3-80 keV) and hard (3-100 keV) X-rays \citep{agrawal2006broad}. It has four science payloads: 1) Soft X-ray Telescope (SXT) \citep[][]{singh2016orbit,singh2017soft}, 2) Ultra-Violet Imaging Telescope (UVIT) \citep[][]{tandon2017orbit}, 3) Cadmium Zinc Telluride Imager (CZTI) \citep[][]{bhalerao2017cadmium} and 4) the Large Area X-ray Proportional Counter (LAXPC) \citep[][]{yadav2016astrosat, antia2017calibration}. Large Area X-ray Proportional Counters (LAXPC) consist of three identical but independent PCUs (LAXPC 10, LAXPC20 and LAXPC30) with an effective area of 6000 cm$^2$ at 15 keV and has a time resolution of 10$\mu$s in the energy range 3.0-80.0 keV with the dead-time of about 42 $\mu$s \citep[][]{yadav2016astrosat,yadav2016large,agrawal2017large}. \\
A simultaneous fit of SXT data along with LAXPC data provides a broadband spectrum of the source. We have analysed various observations with simultaneous data from SXT and LAXPC spanning over 1094 days starting from 3 March 2016. Out of all the AstroSat observations that we looked into, we picked out the ones that showed the presence of QPOs in their power density spectrum. 
In our study, we have included only those observations when the source flux is more or less steady. GRS 1916+105 often shows strong flares when the flux can change by a factor of a few \citep[][]{belloni1997unified, yadav1999different}. Such flaring situations are not included in this study.
\begin{figure*}
     \centering
     \begin{subfigure}[b]{0.49\textwidth}
         \centering
         \includegraphics[width=\textwidth]{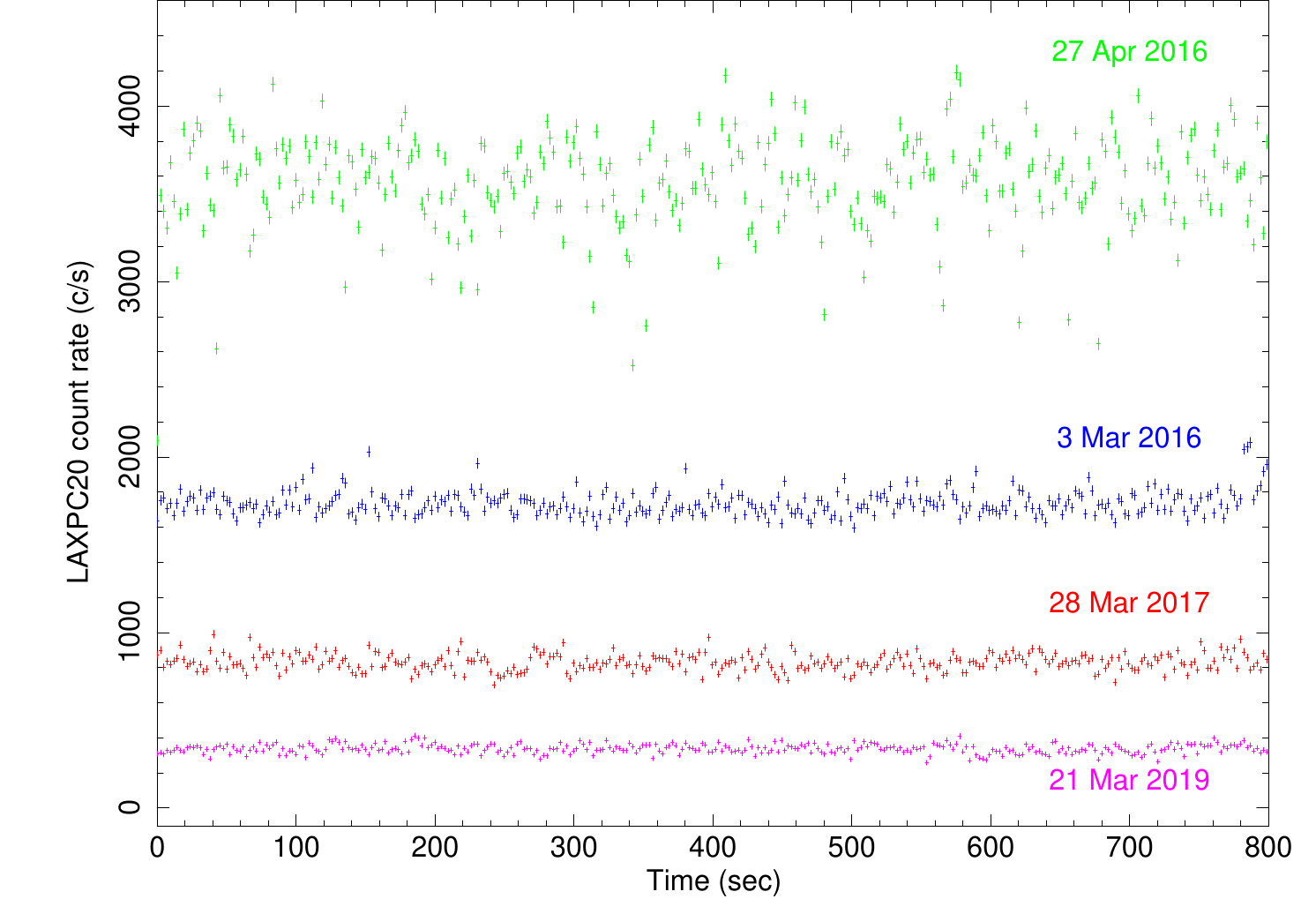}
         \caption{}
          
     \end{subfigure}
     \hfill
     \begin{subfigure}[b]{0.48\textwidth}
         \centering
         \includegraphics[width=\textwidth]{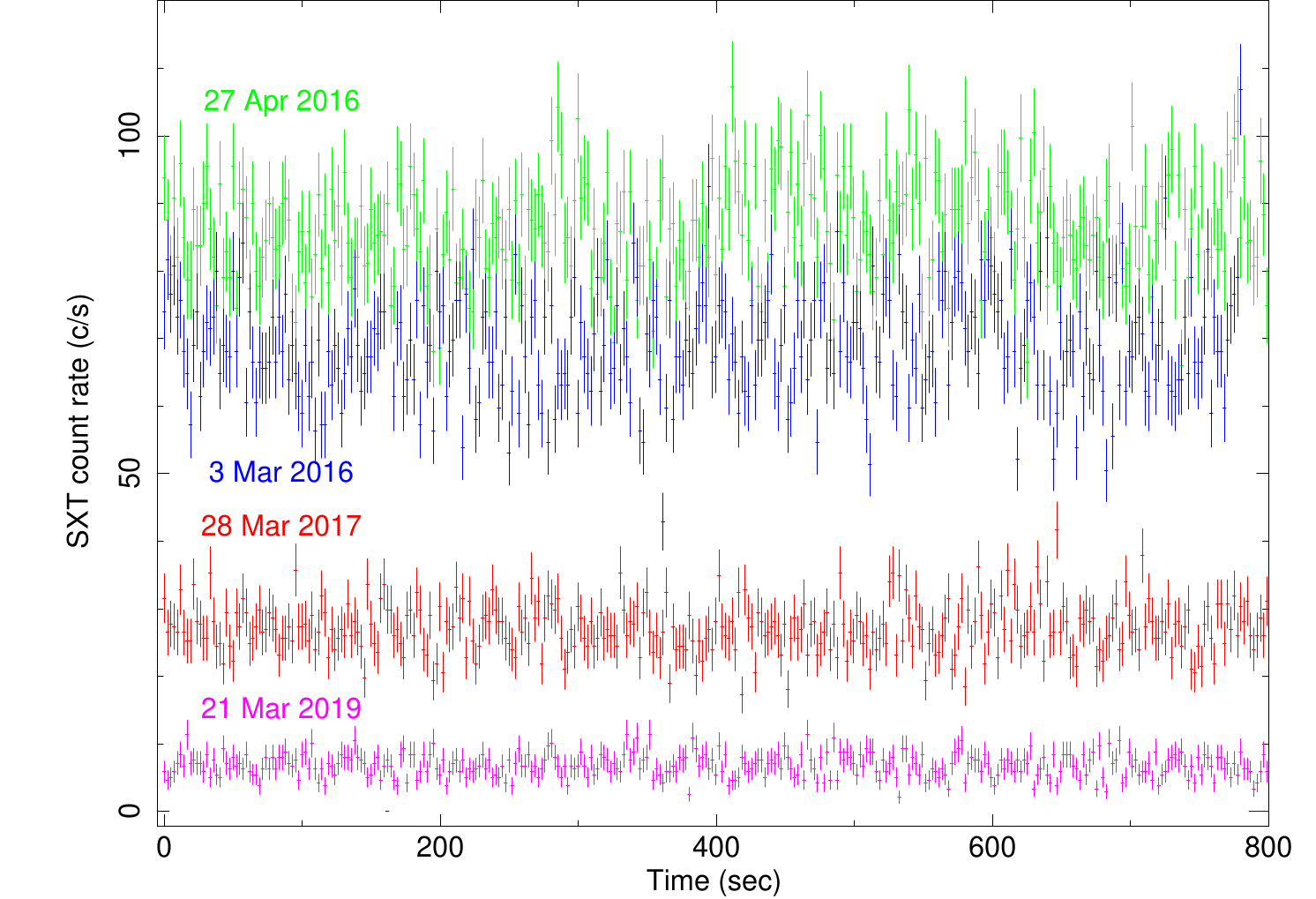}
         \caption{}
          
     \end{subfigure}
    \caption{An 800-sec background-subtracted light curve generated from LAXPC 20 for the observations made on 27 Apr 2016 (Obs. 2), 3 Mar 2016 (Obs. 1), 28 Mar 2017 (Obs. 4), and 21 Mar 2019 (Obs. 7) are shown in the left panel of the figure in four different colours. The right panel represents the SXT light curves for the same observations.}
\label{fig:fig 3}
\end{figure*}
All transient black hole binary outbursts should follow a q-diagram. GRS 1915+105 has shown only one outburst so far; starting with its discovery on 15th August 1992 and the outburst ending now (not yet over); for approximately 31 years.  The rising phase of the outburst in GRS 1915+105 is never observed. Our observations cover the period from 2016 to 2019 when the source remained mostly in luminous  X-ray states. Thus our observations trace only part of the q-diagram; mostly vertical left and bottom horizontal branches, partly when QPOs are present. Its variability is complex as the source stays in the high luminous X-ray states most of the time. We selected seven observations of four distinct states: the High Soft (HS) state, the Low HIMS state; the High HIMS state; and the Low Hard (LS) state.

The data used in this work consists of 7 different observations made on  3 March 2016 (Obs. 1), 25 April 2016 (Obs. 2), 27 April 2016 (Obs. 3), 28 March 2017 (Obs. 4), 1 April 2017 (Obs. 5), 15 April 2017 (Obs. 6), and 21 March 2019 (Obs. 7). Table \ref{tab:table1} presents the effective exposure time of LAXPC and SXT of the observations used in this study. The Burst Alert Telescope (SWIFT/BAT) Hard X-ray Transient Monitor and the Monitor of All-sky X-ray Imaging (MAXI) provide continuous coverage of GRS 1915+105 in soft and hard X-rays. To see the evolution of the source, we extract the MAXI flux in the energy range of 2–20 keV and the SWIFT/BAT flux in the energy range of 15–50 keV, as shown in Fig. \ref{fig:fig1}. The SWIFT/BAT flux is scaled by 30 so that both X-ray band light curves of GRS 1915+105 starting from 13 January 2016 to 27 April 2019 can be seen clearly. The vertical lines in the figure represent AstroSat observations of the GRS 1915+105 source used for this study. The sequence of vertical lines in the light curve shown in Fig. \ref{fig:fig1}
is identical to that presented in Table \ref{tab:table1}. Each observation was further divided into segments such that each segment was continuous without gaps. \\
The HID of GRS 1915+105, covering the period from 13 January 2016 (MJD 57400) to 27 April 2019 (MJD 58600), is illustrated in Fig.
\ref{fig:fig2}, where the 2–20 keV MAXI flux is plotted against the X-ray colour (HR). The location of the source in the HID diagram broadly reflects the state of the system. Also marked in Fig. \ref{fig:fig2} are the locations of the AstroSat observations.  Obs. 2 and  3 correspond to the soft state, while the high flux of Obs. 1 shows that it is in the Hard Intermediate state (High HIMS). On the other hand, Obs. 4,  5 and  6 correspond to
the Low HIMS state. The data from Obs. 7 represents the Low Hard
state of the source. 

\begin{figure*}
     \centering
     \begin{subfigure}[b]{0.49\textwidth}
         \centering
         \includegraphics[width=\textwidth]{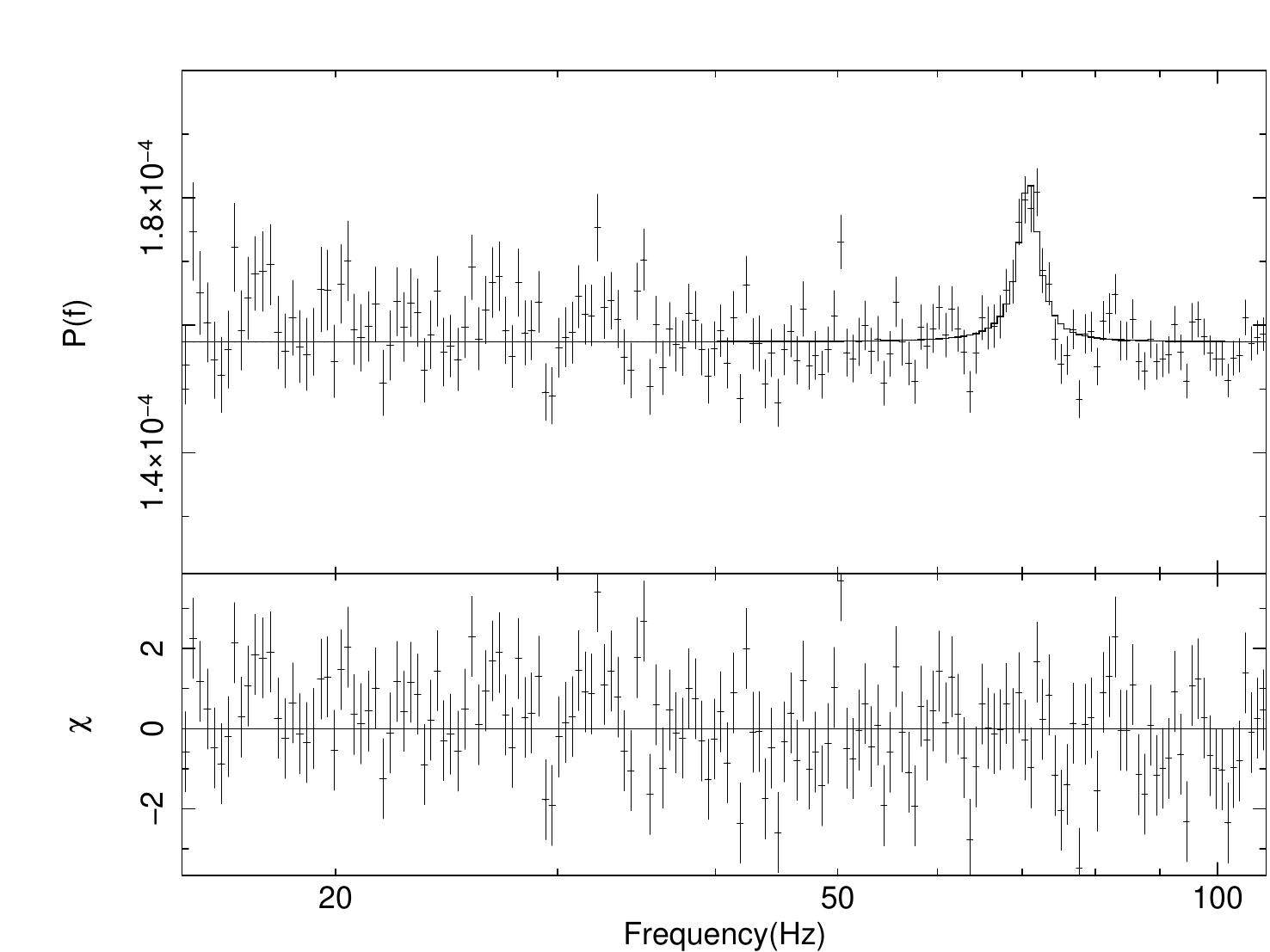}
         \caption{}
          
     \end{subfigure}
     \hfill
     \begin{subfigure}[b]{0.49\textwidth}
         \centering 

         \includegraphics[width=\textwidth]{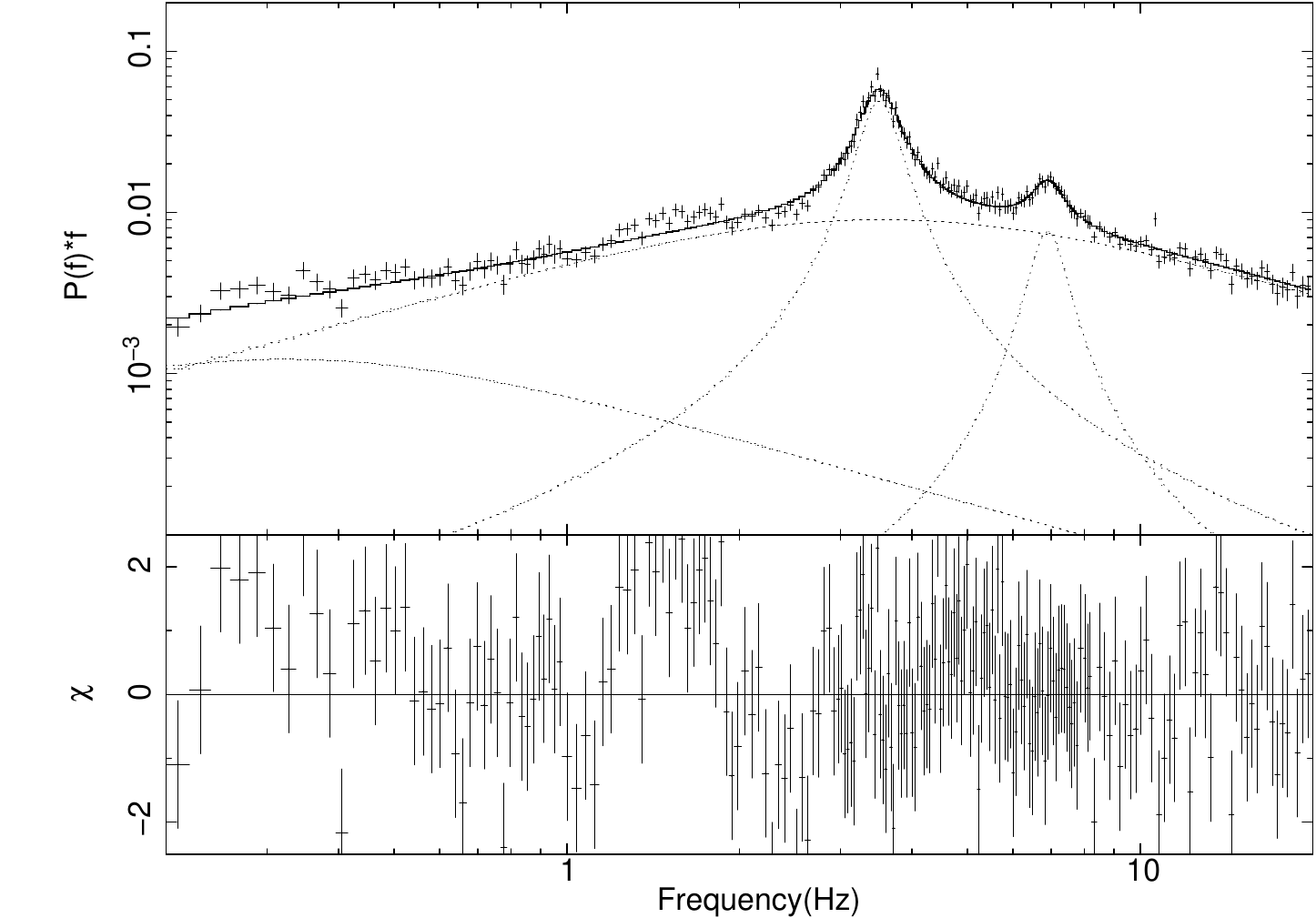}
         \caption{}
          
     \end{subfigure}
     \hfill
     \hfill
     \begin{subfigure}[b]{0.49\textwidth}
         \centering
         \includegraphics[width=\textwidth]{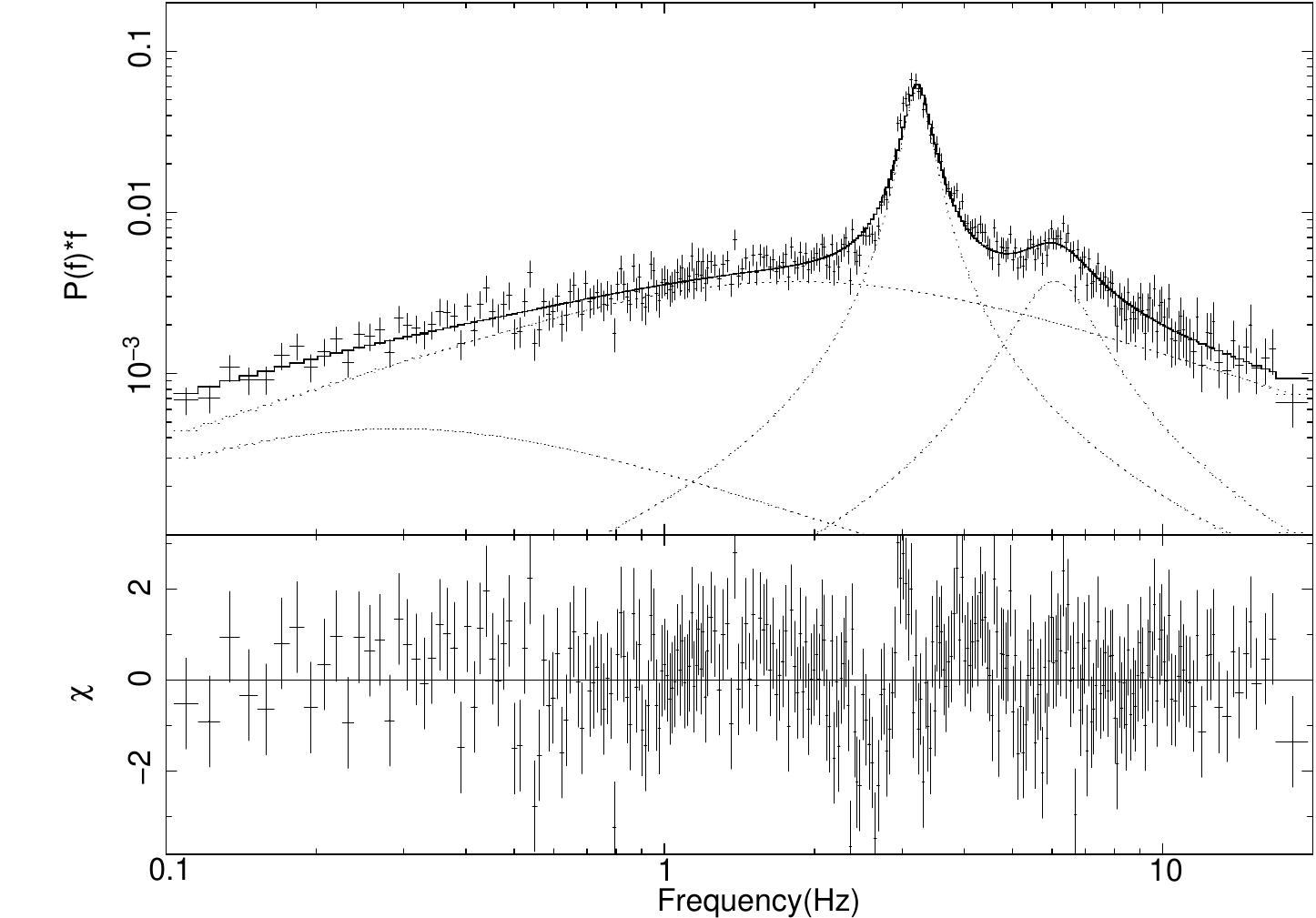}
         \caption{}
          
     \end{subfigure}
     \hfill
     \begin{subfigure}[b]{0.505\textwidth}
         \centering
         \includegraphics[width=\textwidth]{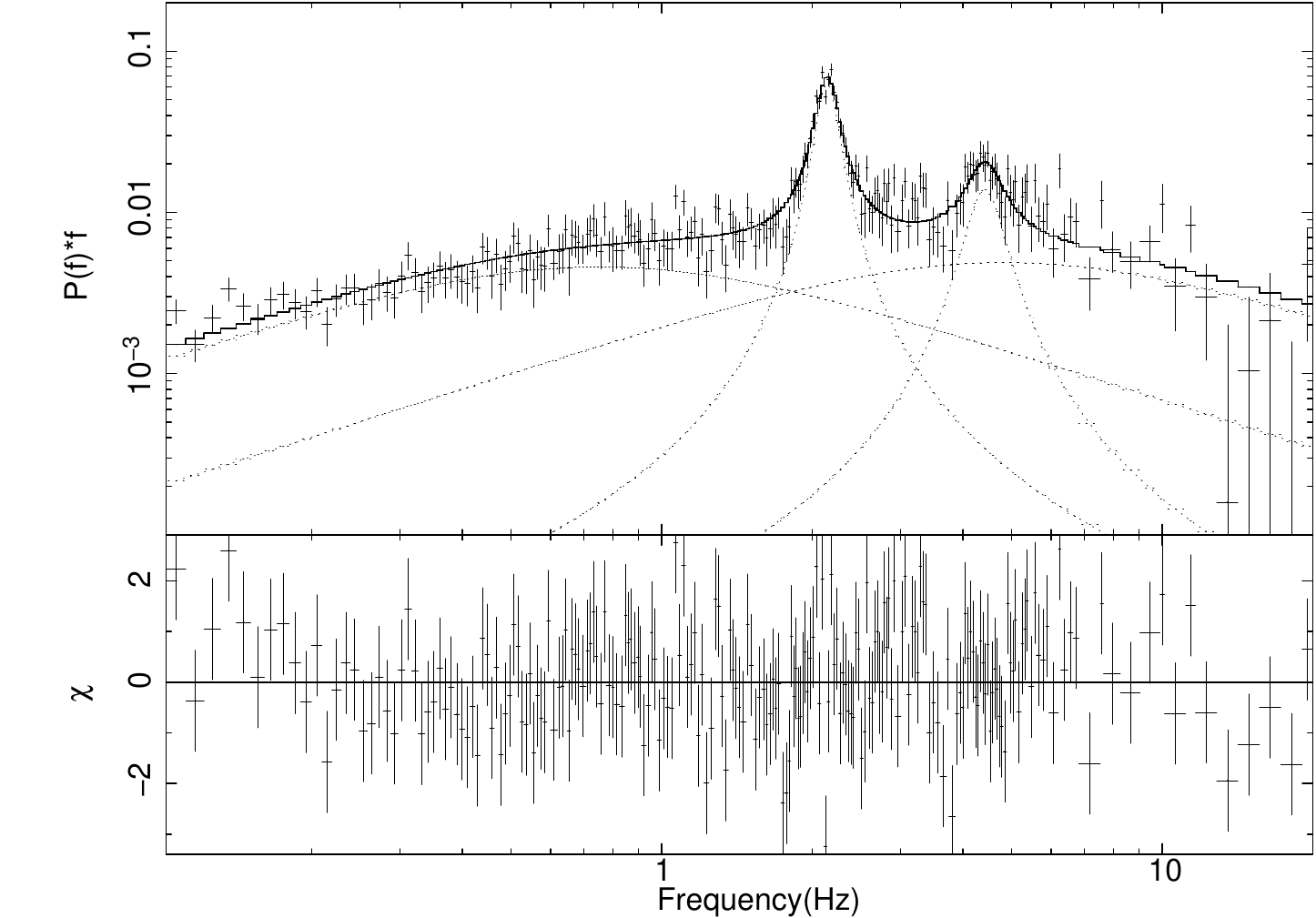}
         \caption{}
          
     \end{subfigure}
     \hfill
        \caption{The top left panel shows the power density spectra of the HS state (Obs. 3, Seg. 2) in the 10-110 Hz frequency range, utilising all three LAXPC detectors. The top right panel shows the PDS of low HIMS state (Obs. 4, Seg. 6) in the frequency range of 0.1-20 Hz. The bottom left panels show the PDS in the frequency range 0.1-20 Hz for observations of SPL state (Obs 1, Seg. 5) and LH state (Obs 7, Seg. 7) in the bottom right panel. LAXPC 20 (only single detector) data is used for these analyses.}
        \label{fig:fig4}
\end{figure*}

\subsection{SXT Data Reduction}

Level 1 photon counting mode data of the SXT instrument was processed through the official
SXT pipeline AS1SXTLevel2 - 1.4b\footnote{\href{https://www.tifr.res.in/~astrosat_sxt/sxtpipeline.html}{https://www.tifr.res.in/~astrosat\_sxt/sxtpipeline.html}} to produce Level 2 mode data. The Photon
Counting mode (PC mode) data were chosen for the analysis of all sets of observations 
listed in Table \ref{tab:table1}.
Using Julia-based SXTevtmerger script\footnote{\label{ft3}\href{https://www.tifr.res.in/~astrosat_sxt/dataanalysis.html}{https://www.tifr.res.in/~astrosat\_sxt/dataanalysis.html}}, we merged all the events belonging to
one set of observations into a single event file. The HEASoft (version 6.29) tool XSELECT was used to generate the spectrum, light curves and images. The response matrix file (RMF) “sxt\_pc\_mat\_g0to12\_RM.rmf,” standard background spectrum “SkyBkg\_comb\_EL3p5\_Cl\_Rd16p0\_v01.pha” and ancillary response file (ARF) "sxt\_pc\_excl00\_v04\_20190608\_mod\_16oct21.arf" were used for the analysis. The sxtARFmodule\footnote{\href{https://www.tifr.res.in/~astrosat_sxt/sxtpipeline.html}{https://www.tifr.res.in/~astrosat\_sxt/sxtpipeline.html}} provided by the SXT
instrument team was used to apply a correction for offset pointing. In order to implement simultaneous analysis, we ensured that the LAXPC 20 observations were available at the same Good Time Interval (GTI) as the SXT observation. Therefore, we
used the simultaneous data segments to generate light curves, images and spectrum of
GRS1915+105.

For the Obs. 4, Obs. 5, Obs. 6 and Obs. 7 observations (low X-ray flux states), there was no pile-up near the centre of the image due to low flux (<40 counts per second, as mentioned in the AstroSat Handbook;\footnote{\href{https://www.iucaa.in/~astrosat/AstroSat_handbook.pdf}{https://www.iucaa.in/~astrosat/AstroSat\_handbook.pdf}}). The average count rate in the Obs. 1, Obs. 2 and Obs. 3 was 91.33 counts/sec, 84.25 counts/sec, and 90.00 counts/sec, respectively. Therefore, to account for the pile-up effect at the centre of the image caused by the high flux rate ($\sim$ 1 Crab) of the source in the charged-coupled device (CCD), the inner radius of the circular annulus region was set to 2 arcmins.

\subsection{LAXPC Data Reduction}
\label{subsec:laxpc}
Level 2 event files were extracted from Level 1 event mode data utilising the official LAXPC software version released on 04 Aug 2020\footnote{\href{http://astrosat-ssc.iucaa.in/laxpcData}{http://astrosat-ssc.iucaa.in/laxpcData}}.
LAXPC data was extracted to obtain the light curve and spectrum of the source\citep{agrawal2018spectral,sreehari2019astrosat}.
Details of the response matrix (RMF) and background spectrum generation for
proportional counters 10, 20, and 30, respectively, can be found in \citet{antia2017calibration}.

Out of three LAXPC detectors (LAXPC 10, LAXPC20 and LAXPC 30), we used only LAXPC 20 
data for energy spectral studies for all of the observations given in Table \ref{tab:table1}.
\section{Data Analysis}
\label{sec:data_analysis}

\subsection{X-ray lighcurve and Timing Analysis}
\label{subsec:timing}
\begin{figure*}
     \centering
     \begin{subfigure}[b]{0.49\textwidth}
         \centering
         \includegraphics[width=\textwidth]{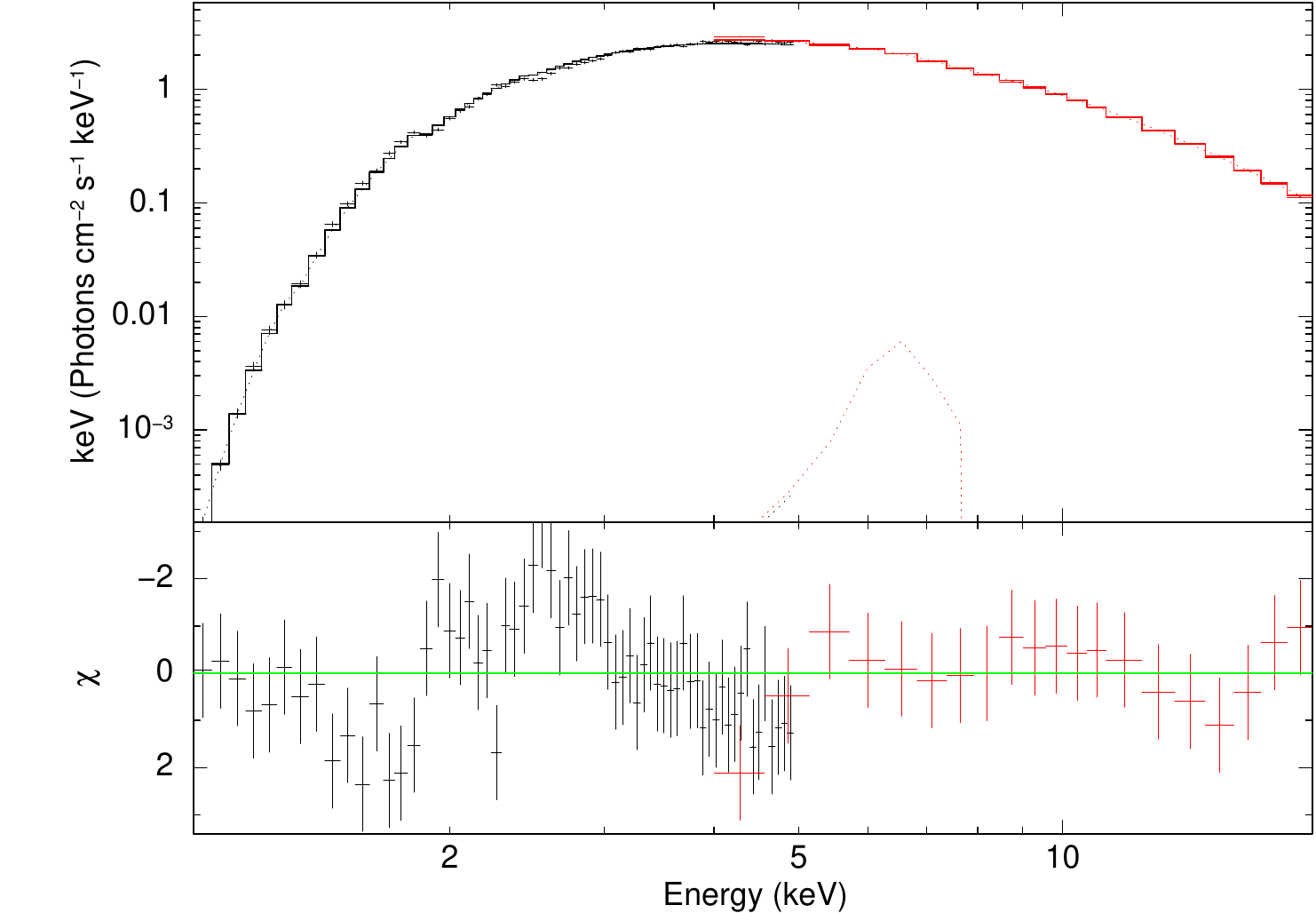}
         \caption{}
          
     \end{subfigure}
     \hfill
     \begin{subfigure}[b]{0.495\textwidth}
         \centering
         \includegraphics[width=\textwidth]{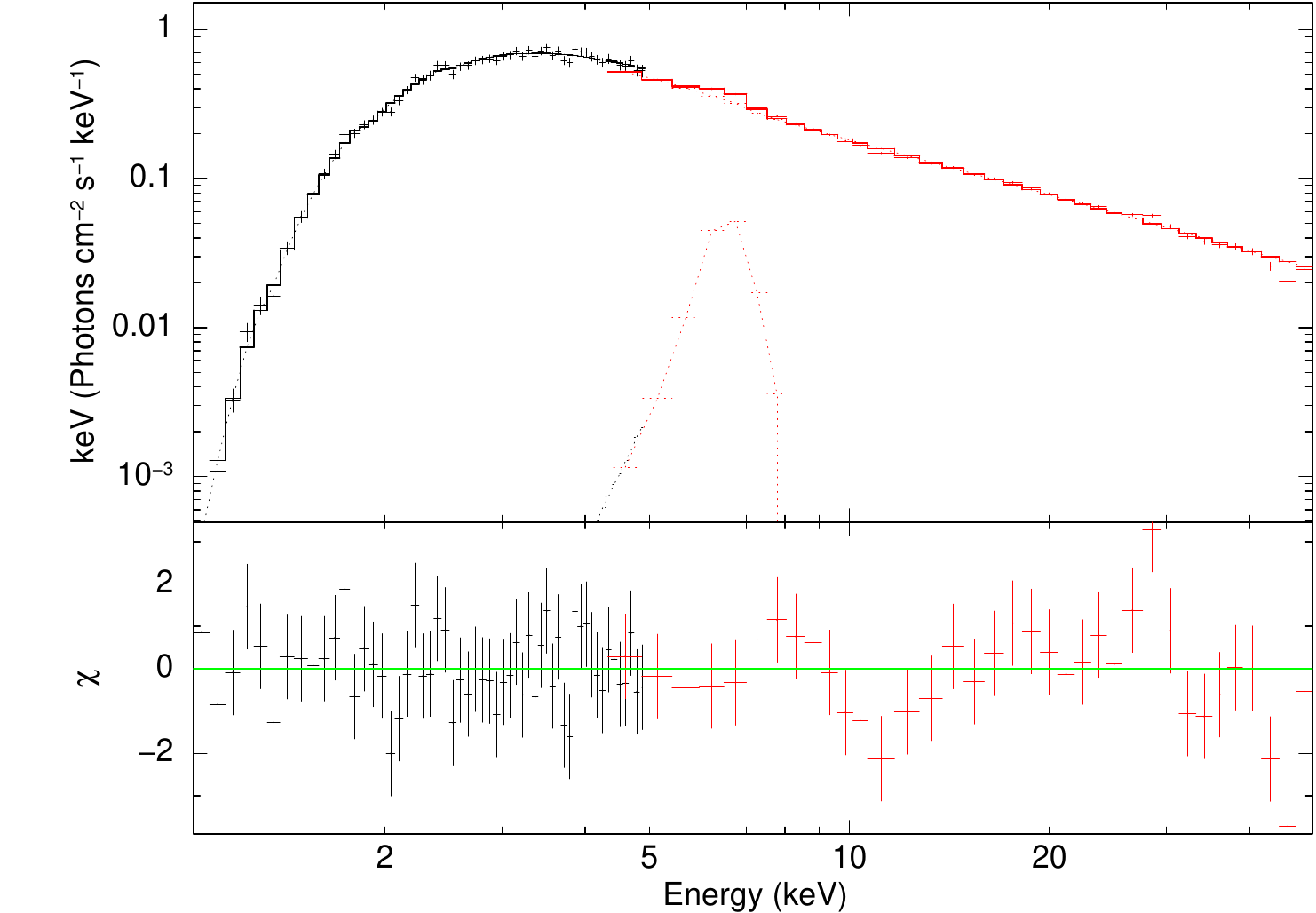}
         \caption{}
          
     \end{subfigure}
     \hfill
     \vspace{10pt}
     \begin{subfigure}[b]{0.49\textwidth}
         \centering
     \includegraphics[width=\textwidth]{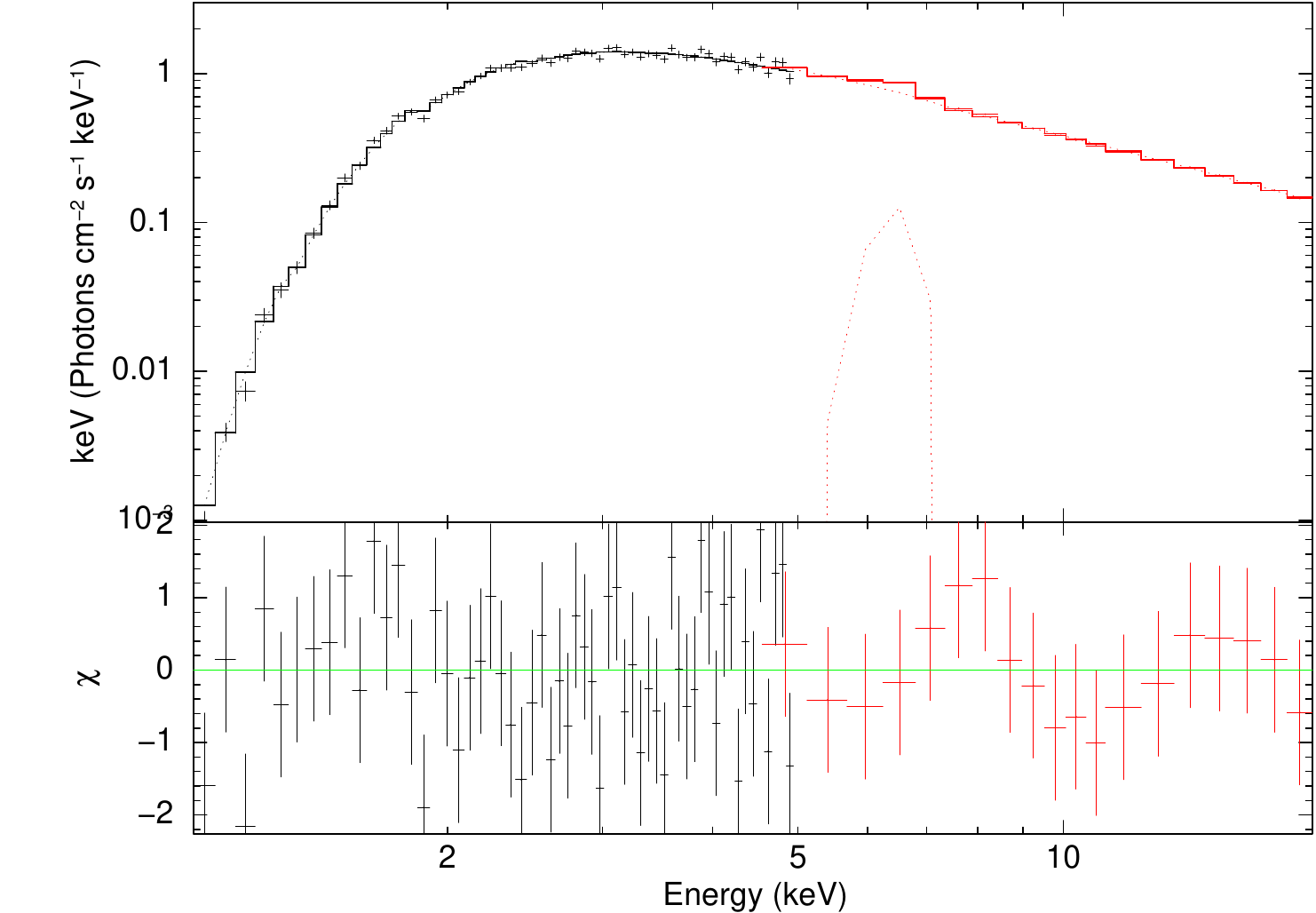}
         \caption{}
          
     \end{subfigure}
       \hfill
     \begin{subfigure}[b]{0.49\textwidth}
         \centering
         \includegraphics[width=\textwidth]{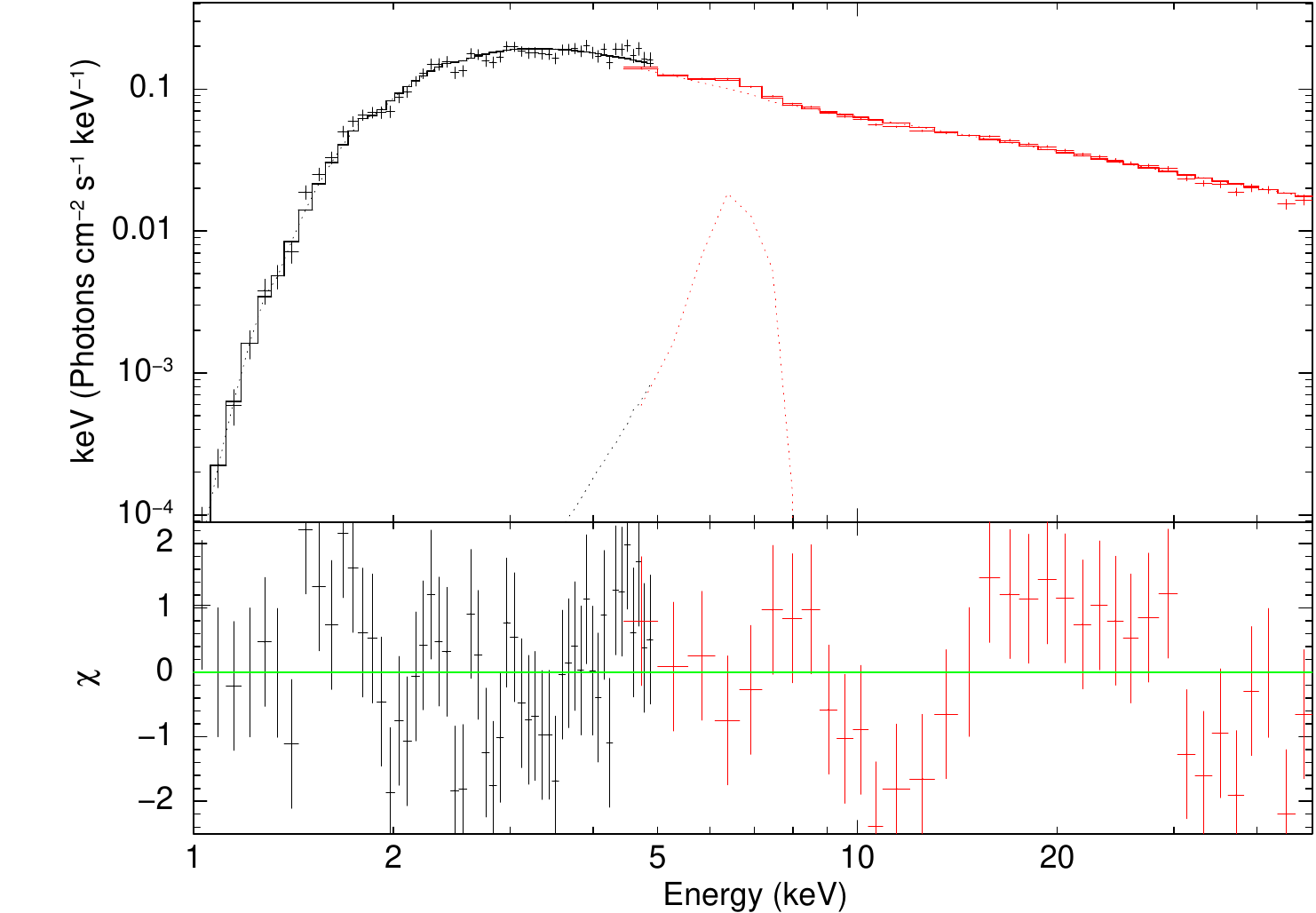}
         \caption{}
          
     \end{subfigure}
        \caption{The left panels show the energy spectra of  HS state (Obs. 3, Seg. 2) in the top panel and SPL state (Obs. 1, Seg. 5 ) in the bottom panel,  respectively, in the energy range of 1-20 keV. The low HIMS and LH state spectra for the Obs. 4 (Seg. 6) and Obs. 7 (Seg. 7), respectively, are shown in the right top and right bottom  panels in the energy range of 1-50 keV. The residuals  are displayed beneath each spectrum. The SXT and LAXPC data points are black and red, respectively.}
        \label{fig:fig5}
\end{figure*}
We have produced Background-subtracted light curves for four distinct observation types in the 4.0-50 keV energy range using LAXPC 20 data for the minimum time resolution of the SXT, which is 2.378 seconds. The left panel of Fig. \ref{fig:fig 3} shows  800 sec long Background-Subtracted light curve for HS state (Obs. 3), SPL state (Obs. 1), low HIMS (Obs. 4), and LH state (Obs 7).
The right panel of Fig.\ref{fig:fig 3} shows 800 sec SXT light curves in the 0.3-8 keV energy range for the identical segments used to generate the LAXPC 20 Background-Subtracted lightcurves in the left panel.

In order to study the properties of QPOs, we analyse the data in the frequency regime by generating a Power Density Spectrum (PDS). The PDS were generated by dividing the lightcurve of each segment into parts and averaging the power spectra of each part.   We used all three LAXPC detector units (LAXPC 10, LAXPC 20, and LAXPC 30) to plot the PDS for the HS state (Obs. 3, Seg. 2). To plot the PDS for the rest of the observations, we used the LAXPC 20 unit. The PDS for the HS state is shown in the upper left panel of Fig. \ref{fig:fig4} in the frequency range 10-110 Hz and is modelled using several Lorentzian functions \citep[][]{belloni2002unified, nowak2000there} and a
power-law component in order to account very low frequency noise (VLFN). It shows HFQPO at $\sim$ 70 Hz, while no QPO is seen in the lower frequency region. 

Fig. \ref{fig:fig4}, the upper right panel, shows the PDS of  the  low HIMS state (Obs. 4, Seg. 6) in the frequency range 0.1-20 Hz. The lower  panels of Fig.  \ref{fig:fig4}  show the PDS for the SPL  state  (left panel) and the LH state (right panel) for the Obs. 1 (Seg. 5) and Obs. 7 (Seg. 7), respectively.
The component of broad-band noise related to these three PDS (Obs. 4, Obs. 1, and Obs. 7) was modelled using only a few Lorentzians. The frequency of QPOs, along with errors, has been estimated and tabulated in the third column of Table \ref{tab:table2}. All three panels show LFQPOs along with their harmonics.
\begin{table*}
\caption{Best fit spectral parameters of GRS 1915+105.}
\label{tab:table2}
\begin{threeparttable}
\begin{center}
\begin{tabular}{M{0.8cm} M{0.5cm} M{0.7cm} M{1.1cm} M{1.4cm} M{1.3cm} M{1.6cm} M{1.4cm} M{1.5cm} M{1.7cm} M{1.6cm}}
\hline \hline
Exposure Time LXP20 (sec) & Exposure Time SXT (sec) & Segment & QPO Freq. (Hz) & nH ($10^{22} \textrm{cm}^{-2}$) & Accretion Rate ($10^{18} \textrm{gm/s}$) & Inner Radius (Rg) & Fraction Scatter & Gamma & Flux in line emission ($10^{-2}\textrm{photons}$ $  \textrm{cm}^{-2}  \textrm{s}{-1}$) & $\chi^2/$Dof \\[5pt]
\hline \\

\multicolumn{10}{c}{ SPL/HIMS (Obs.1)} \\[5pt]
1334 & 813 & 1 & $4.51^{+0.02}_{-0.02}$ & 4.00\tnote{a} & $4.51^{+0.43}_{-0.38}$ & $10.60^{+0.70}_{-0.77}$ & 0.6\tnote{a} & $2.874^{+0.017}_{-0.018}$ & $<1.978$ & 67.46/89 \\ [6pt]
\hline \\
1757 & 815 & 2 & $3.54^{+0.02}_{-0.02}$ & 4.00\tnote{a} & $5.22^{+0.50}_{-0.45}$ & $14.00^{+1.25}_{-1.16}$ & 0.6\tnote{a} & $2.818^{+0.014}_{-0.018}$ & $<1.186$ & 100.61/89 \\ [6pt]
\hline \\
2183 & 815 & 3 & $2.56^{+0.01}_{-0.01}$ & 4.00\tnote{a} & $4.18^{+0.47}_{-0.41}$ & $15.53^{+1.73}_{-1.48}$ & 0.6\tnote{a} & $2.571^{+0.036}_{-0.036}$ & $2.155^{+1.425}_{-1.159}$ & 47.20/72 \\ [6pt]
\hline \\
2606 & 803 & 4 & $2.79^{+0.01}_{-0.01}$ & 4.00\tnote{a} & $4.91^{+0.52}_{-0.52}$ & $18.16^{+1.58}_{-2.15}$ & 0.6\tnote{a} & $2.586^{+0.036}_{-0.036}$ &  $2.446^{+1.313}_{-0.980}$ & 73.23/72 \\ [6pt]
\hline \\
3034 & 815 & 5 & $3.22^{+0.01}_{-0.01}$ & 4.00\tnote{a} & $4.55^{+0.51}_{-0.22}$ & $15.33^{+1.47}_{-1.40}$ & 0.6\tnote{a} & $2.630^{+0.037}_{-0.057}$ &  $2.374^{+1.271}_{-1.230}$ & 48.66/72 \\ [6pt]
\hline \\
354 & 303 & 6 & $3.82^{+0.04}_{-0.04}$ & 4.00\tnote{a} & $4.28^{+0.66}_{-0.51}$ & $12.58^{+1.43}_{-1.46}$ & 0.6\tnote{a} & $2.715^{+0.039}_{-0.059}$ & $2.832^{+1.465}_{-1.411}$ & 61.65/70 \\ [5pt]
\hline \\
3535 & 813 & 7 & $4.77^{+0.02}_{-0.02}$ & 4.00\tnote{a} & $5.25^{+0.48}_{-0.48}$ & $11.79^{+0.84}_{-0.88}$ & 0.6\tnote{a} & $2.714^{+0.039}_{-0.039}$ & $2.485^{+1.854}_{-1.827}$ & 56.69/73 \\ [5pt]
\hline \\
3378 & 668 & 8 & $4.94^{+0.02}_{-0.02}$ & 4.00\tnote{a} & $4.17^{+0.41}_{-0.40}$ & $8.96^{+0.65}_{-0.79}$ & 0.6\tnote{a} & $2.816^{+0.041}_{-0.041}$ & $1.222^{+2.041}_{-2.008}$ & 63.22/73 \\ [5pt] 
\hline \\
\multicolumn{10}{c}{HS State (Obs. 2 and Obs. 3)} \\[5pt]
499 & 506 & 9 & $70.54^{+0.41}_{-0.53}$ & $4.65^{+0.12}_{-0.12}$ & $1.69^{+0.01}_{-0.01}$ & $<2.78$ & $0.410^{+0.033}_{-0.030}$ & 4.5\tnote{a} & 1.0\tnote{a} & 70.35/72 \\ [6pt]
\hline \\
1640 & 1640 & 10 & $69.81^{+0.83}_{-0.81}$ & $4.51^{+0.05}_{-0.05}$ & $1.97^{+0.07}_{-0.05}$ & $<1.79$ & $0.351^{+0.031}_{-0.031}$ & 4.5\tnote{a} & 1.0\tnote{a} & 111.69/81 \\ [5pt]
\hline \\
1241 & 1241 & 11 & $71.43^{+0.96}_{-0.95}$ & $4.57^{+0.04}_{-0.05}$ & $2.00^{+0.06}_{-0.02}$ & $<1.81$ & $0.402^{+0.025}_{-0.031}$ & 4.5\tnote{a} & 1.0\tnote{a} & 102.2/79 \\ [5pt]
\hline \\
1077 & 1079 & 12 & $70.79^{+0.84}_{-0.82}$ & $4.47^{+0.06}_{-0.06}$ & $1.87^{+0.08}_{-0.03}$ & $<1.88$ & $0.465^{+0.51}_{-0.037}$ & 4.5\tnote{a} & 1.0\tnote{a} & 86.63/78 \\ [5pt]
\hline \\

\multicolumn{10}{c}{ Low HIMS (Obs. 4)} \\[5pt]
1213 & 1224 & 13 & $3.58^{+0.02}_{-0.02}$ & 4.00 & $0.75^{+0.08}_{-0.07}$ & $5.53^{+0.60}_{-0.46}$ & $0.406^{+0.050}_{-0.056}$ & $2.17^{+0.02}_{-0.02}$ & $1.285^{+0.514}_{-0.500}$ & 96.84/89  \\ [6pt]
\hline \\
1213 & 1224 & 14 & $3.45^{+0.02}_{-0.01}$ & 4.00 & $0.85^{+0.08}_{-0.08}$ & $6.42^{+0.84}_{-0.56}$ & $0.416^{+0.060}_{-0.037}$ & $2.17^{+0.02}_{-0.02}$ & $1.266^{+0.510}_{-0.488}$ & 101.51/89  \\ [5pt]
\hline \\
1213 & 1227 & 15 & $3.65^{+0.02}_{-0.02}$ & 4.00 & $0.86^{+0.06}_{-0.07}$ & $6.60^{+0.71}_{-0.75}$ & $0.444^{+0.050}_{-0.055}$ & $2.22^{+0.02}_{-0.03}$ & $1.344^{+0.492}_{-0.482}$ & 84.77/89  \\ [5pt]
\hline \\
951 & 955 & 16 & $4.05^{+0.03}_{-0.02}$ & 4.00 & $0.76^{+0.08}_{-0.08}$ & $5.14^{+0.44}_{-0.48}$ & $0.397^{+0.045}_{-0.039}$ & $2.24^{+0.02}_{-0.03}$ & $1.218^{+0.553}_{-0.565}$ & 90.12/89  \\ [5pt]
\hline \\
523 & 530 & 17 & $4.14^{+0.03}_{-0.03}$ & 4.00 & $0.76^{+0.10}_{-0.08}$ & $5.24^{+0.60}_{-0.55}$ & $0.400^{+0.048}_{-0.040}$ & $2.26^{+0.03}_{-0.03}$ & $1.132^{+0.599}_{-0.484}$ & 89.06/86  \\ [5pt]
\hline \\
1213 & 1232 & 18 & $4.35^{+0.04}_{-0.04}$ & 4.00 & $0.67^{+0.07}_{-0.06}$ & $3.97^{+0.35}_{-0.43}$ & $0.337^{+0.035}_{-0.029}$ & $2.24^{+0.03}_{-0.03}$ & $0.940^{+0.617}_{-0.596}$ & 62.41/90  \\ [5pt]
\hline \\
1209 & 1210 & 19 & $5.09^{+0.03}_{-0.04}$ & 4.00 & $0.68^{+0.07}_{-0.06}$ & $3.58^{+0.30}_{-0.29}$ & $0.284^{+0.027}_{-0.023}$ & $2.25^{+0.03}_{-0.03}$ & $0.748^{+0.645}_{-0.626}$ & 67.42/90  \\ [5pt]
\hline \\
1212 & 1213 & 20 & $5.18^{+0.03}_{-0.03}$ & 4.00 & $0.68^{+0.06}_{-0.06}$ & $3.64^{+0.30}_{-0.30}$ & $0.295^{+0.027}_{-0.025}$ & $2.26^{+0.03}_{-0.03}$ & $0.864^{+0.635}_{-0.632}$ & 90.24/90  \\ [5pt]
\hline \\
1212 & 1213 & 21 & $5.34^{+0.01}_{-0.02}$ & 4.00 & $0.66^{+0.06}_{-0.06}$ & $3.31^{+0.21}_{-0.28}$ & $0.279^{+0.025}_{-0.023}$ & $2.26^{+0.03}_{-0.03}$ & $0.652^{+0.717}_{-0.610}$ & 92.71/90  \\ [5pt]
\hline \\
1216 & 1217 & 22 & $5.41^{+0.02}_{-0.03}$ & 4.00 & $0.72^{+0.07}_{-0.07}$ & $3.56^{+0.02}_{-0.02}$ & $0.259^{+0.024}_{-0.021}$ & $2.26^{+0.03}_{-0.02}$ & $0.796^{+0.648}_{-0.645}$ & 95.11/90  \\ [5pt]
\hline \\

\end{tabular}
\begin{tablenotes}
    \item[a] Parameter frozen during the fitting.
\end{tablenotes}
\end{center}
\end{threeparttable}
\end{table*}

\begin{table*}
\begin{threeparttable}
\begin{center}
\begin{tabular}{M{0.8cm} M{0.8cm} M{0.5cm} M{1.0cm} M{1.4cm} M{1.2cm} M{1.5cm} M{1.3cm} M{1.4cm} M{1.0cm} M{1.6cm}}
\hline \hline
Exposure Time LXP20 (sec) & Exposure Time SXT (sec) & Segment & QPO Freq. (Hz) & nH ($10^{22} \textrm{cm}^{-2}$) & Accretion Rate ($10^{18} \textrm{gm/s}$) & Inner Radius (Rg) & Fraction Scatter & Gamma & Flux in line emission ($10^{-2}\textrm{photons}$ $  \textrm{cm}^{-2}  \textrm{s}{-1}$) & $\chi^2/$Dof \\[5pt]
\hline \\

\multicolumn{10}{c}{Low HIMS (Obs. 5)} \\[5pt]
634 & 638 & 23 & $4.04^{+0.01}_{-0.02}$ & 4.00\tnote{a} & $0.86^{+0.09}_{-0.08}$ & $5.86^{+0.63}_{-0.53}$ & $0.397^{+0.045}_{-0.037}$ & $2.24^{+0.02}_{-0.02}$ & $1.290^{+0.498}_{-0.494}$ & 77.59/86  \\ [8pt]
\hline \\
204 & 204 & 24 & $4.67^{+0.02}_{-0.03}$ & 4.00\tnote{a} & $0.84^{+0.16}_{-0.13}$ & $4.94^{+0.78}_{-0.73}$ & $0.349^{+0.042}_{-0.038}$ & $2.27^{+0.03}_{-0.04}$ & $1.068^{+0.752}_{-0.552}$ & 88.08/81  \\ [8pt]
\hline \\
748 & 746 & 25 & $5.14^{+0.02}_{-0.02}$ & 4.00\tnote{a} & $0.72^{+0.08}_{-0.07}$ & $3.95^{+0.37}_{-0.36}$ & $0.311^{+0.032}_{-0.027}$ & $2.28^{+0.03}_{-0.03}$ & $0.922^{+0.648}_{-0.635}$ & 68.21/88  \\ [8pt]
\hline \\
1175 & 1177 & 26 & $5.13^{+0.03}_{-0.03}$ & 4.00\tnote{a} & $0.70^{+0.07}_{-0.06}$ & $3.78^{+0.31}_{-0.29}$ & $0.309^{+0.030}_{-0.026}$ & $2.30^{+0.03}_{-0.01}$ & $0.848^{+0.638}_{-0.630}$ & 88.94/90  \\ [8pt]
\hline \\
1260 & 1260 & 27 & $5.10^{+0.03}_{-0.02}$ & 4.00\tnote{a} & $0.75^{+0.07}_{-0.07}$ & $4.16^{+0.31}_{-0.30}$ & $0.307^{+0.030}_{-0.027}$ & $2.29^{+0.03}_{-0.03}$ & $0.886^{+0.615}_{-0.606}$ & 99.64/90  \\ [8pt]
\hline \\
762 & 763 & 28 & $5.31^{+0.03}_{-0.04}$ & 4.00\tnote{a} & $0.70^{+0.07}_{-0.07}$ & $3.64^{+0.34}_{-0.34}$ & $0.286^{+0.027}_{-0.025}$ & $2.27^{+0.03}_{-0.03}$ & $0.692^{+0.517}_{-0.504}$ & 59.87/87  \\ [8pt]
\hline \\
\multicolumn{10}{c}{Low HIMS (Obs. 6)} \\[5pt]

1465 & 1465 & 29 & $4.52^{+0.03}_{-0.03}$ & 4.00 & $0.77^{+0.07}_{-0.07}$ & $4.70^{+0.31}_{-0.32}$ & $0.302^{+0.029}_{-0.027}$ & $2.30^{+0.01}_{-0.03}$ & $0.692^{+0.5}_{-0.5}$ & 129.81/91  \\ [8pt]
\hline \\
1184 & 1193 & 30 & $4.51^{+0.03}_{-0.03}$ & 4.00 & $0.74^{+0.07}_{-0.06}$ & $4.60^{+0.34}_{-0.32}$ & $0.306^{+0.030}_{-0.026}$ & $2.29^{+0.03}_{-0.03}$ & $0.816^{+0.516}_{-0.505}$ & 81.87/90  \\ [8pt]
\hline \\
1362 & 1369 & 31 & $4.48^{+0.05}_{-0.04}$ & 4.00 & $0.75^{+0.07}_{-0.06}$ & $4.65^{+0.32}_{-0.31}$ & $0.306^{+0.029}_{-0.026}$ & $2.29^{+0.03}_{-0.03}$ & $0.770^{+0.564}_{-0.445}$ & 66.60/90  \\ [8pt]
\hline \\

\multicolumn{10}{c}{LH State (Obs. 7)} \\[5pt]
1082 & 1082 & 32 & $2.42^{+0.01}_{-0.01}$ & 4.00 & $0.15^{+0.03}_{-0.03}$ & $3.13^{+0.61}_{-0.64}$ & $0.325^{+0.040}_{-0.017}$ & $1.84^{+0.02}_{-0.03}$ & $0.522^{+0.189}_{-0.274}$ & 103.9/82  \\ [8pt]
\hline \\
1082 & 1082 & 33 & $2.41^{+0.01}_{-0.01}$ & 4.00 & $0.14^{+0.03}_{-0.02}$ & $2.68^{+0.59}_{-0.81}$ & $0.335^{+0.041}_{-0.034}$ & $1.85^{+0.02}_{-0.03}$ & $0.547^{+0.202}_{-0.191}$ & 81.4/83  \\ [8pt]
\hline \\
1084 & 1084 & 34 & $2.42^{+0.01}_{-0.01}$ & 4.00 & $0.16^{+0.03}_{-0.03}$ & $3.28^{+0.62}_{-0.61}$ & $0.325^{+0.040}_{-0.035}$ & $1.86^{+0.03}_{-0.03}$ & $0.419^{+0.189}_{-0.165}$ & 104.9/83  \\ [8pt]
\hline \\
1091 & 1091 & 35 & $2.14^{+0.02}_{-0.02}$ & 4.00 & $0.15^{+0.03}_{-0.03}$ & $3.11^{+0.61}_{-0.65}$ & $0.331^{+0.042}_{-0.036}$ & $1.79^{+0.02}_{-0.02}$ & $0.712^{+0.306}_{-0.101}$ & 129.2/82  \\[8pt]
\hline \\
1091 & 1091 & 36 & $2.01^{+0.01}_{-0.01}$ & 4.00 & $0.14^{+0.03}_{-0.02}$ & $3.14^{+0.66}_{-0.72}$ & $0.363^{+0.046}_{-0.020}$ & $1.82^{+0.03}_{-0.03}$ & $0.497^{+0.190}_{-0.192}$ & 110.5/83  \\ [8pt]
\hline \\
1096 & 1096 & 37 & $2.14^{+0.01}_{-0.01}$ & 4.00 & $0.16^{+0.03}_{-0.03}$ & $3.58^{+0.73}_{-0.63}$ & $0.340^{+0.044}_{-0.037}$ & $1.83^{+0.03}_{-0.03}$ & $0.540^{+0.180}_{-0.177}$ & 107.5/83  \\ [8pt]
\hline \\
998 & 998 & 38 & $2.04^{+0.02}_{-0.01}$ & 4.00 & $0.13^{+0.03}_{-0.02}$ & $3.03^{+0.90}_{-0.82}$ & $0.392^{+0.058}_{-0.047}$ & $1.86^{+0.03}_{-0.03}$ & $0.484^{+0.193}_{-0.182}$ & 115.5/82  \\ [8pt]
\hline \\
1072 & 1070 & 39 & $2.07^{+0.01}_{-0.01}$ & 4.00 & $0.14^{+0.03}_{-0.02}$ & $2.91^{+0.51}_{-0.67}$ & $0.349^{+0.041}_{-0.037}$ & $1.81^{+0.03}_{-0.03}$ & $0.517^{+0.200}_{-0.188}$ & 125.1/83  \\ [8pt]
\hline \\
1006 & 1006 & 40 & $2.14^{+0.02}_{-0.02}$ & 4.00 & $0.11^{+0.01}_{-0.02}$ & $<3.05$ & $0.376^{+0.041}_{-0.050}$ & $1.82^{+0.02}_{-0.03}$ & $0.476^{+0.152}_{-0.194}$ & 109.3/82  \\ [8pt]
\hline \\
\end{tabular}
\begin{tablenotes}
    \item[a] Parameter frozen during the fitting.
\end{tablenotes}
\end{center}
\end{threeparttable}
\end{table*}

\begin{figure*}
     \centering
     \begin{subfigure}[b]{0.495\textwidth}
         \centering
         \includegraphics[width=\textwidth]{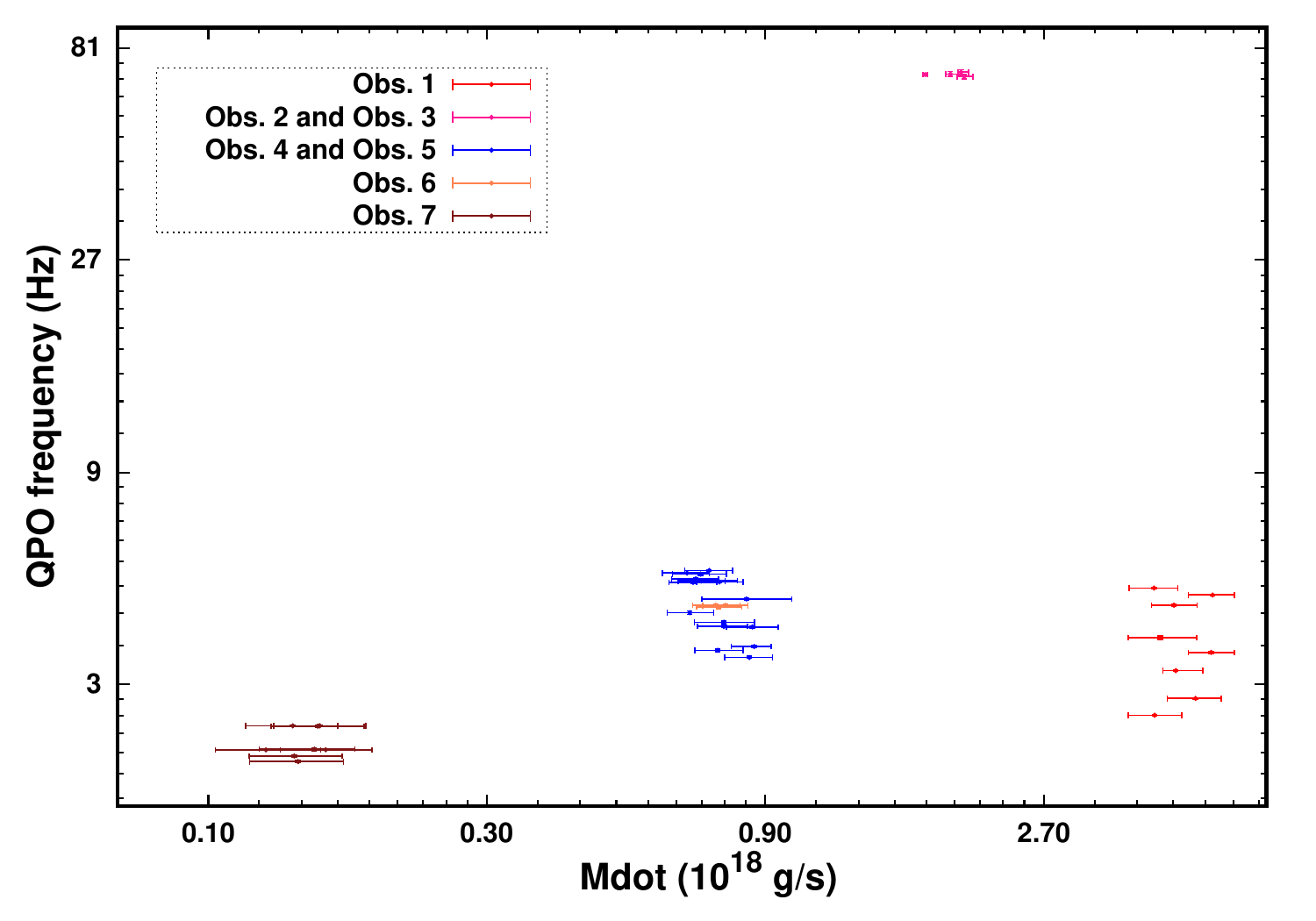}
         \caption{}
          
     \end{subfigure}
     \hfill
     \begin{subfigure}[b]{0.495\textwidth}
         \centering
         \includegraphics[width=\textwidth]{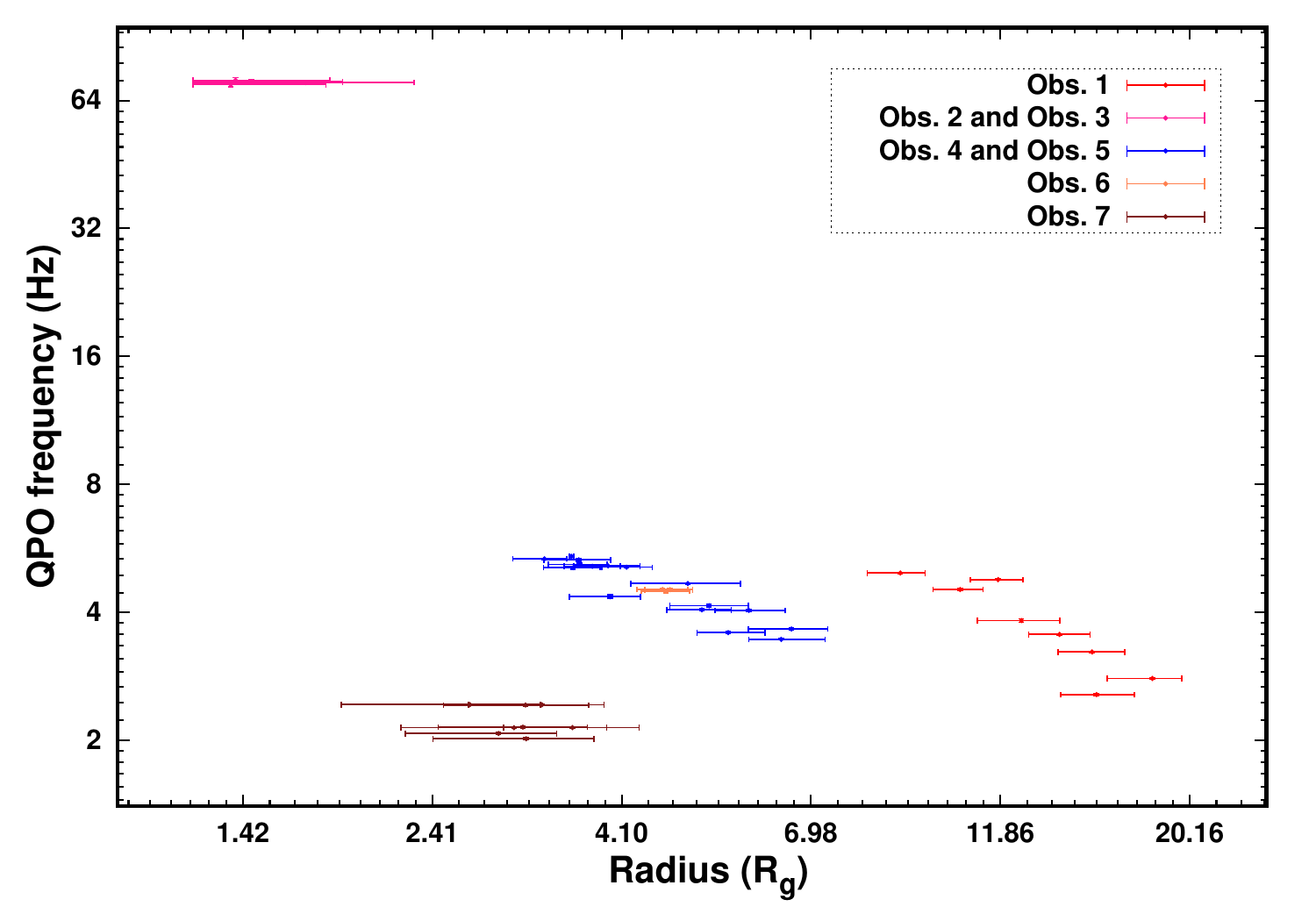}
         \caption{}
          
     \end{subfigure}
     \hfill \\[10pt]
     \begin{subfigure}[b]{0.6\textwidth}
         \centering
         \includegraphics[width=\textwidth]{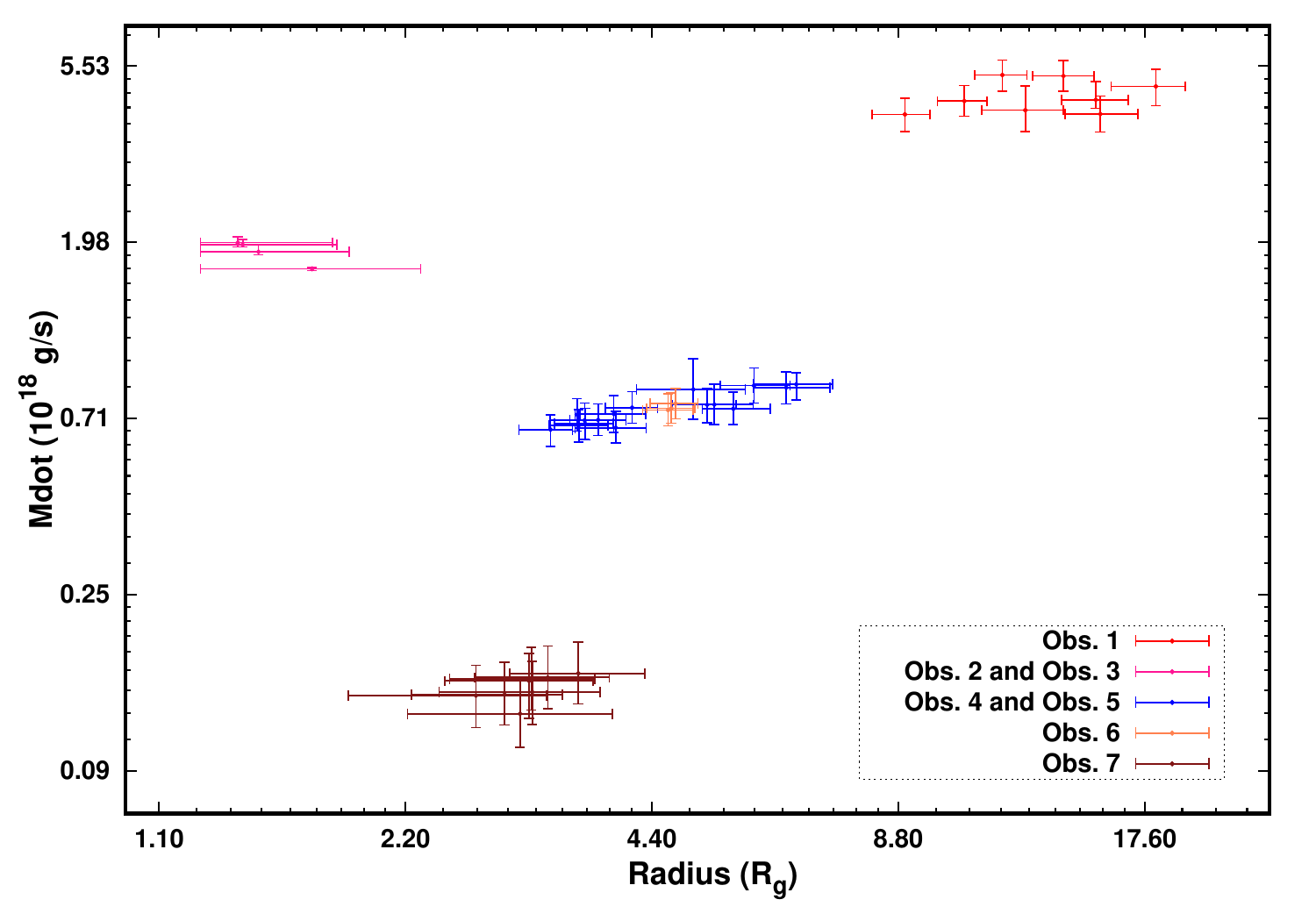}
         \caption{}
          
     \end{subfigure}
        \caption{The top left and top right panels illustrate the variability of QPO frequencies with respect to accretion rate and inner disc radius, respectively. The bottom panel represents the relationship between the accretion rate and the inner disc radius.}
        \label{fig:fig6}
\end{figure*}

\subsection{Spectral Analysis}
\label{subsec:spectral}
We have performed a simultaneous spectral fitting of SXT and LAXPC20 spectra using
XSPEC 12.12.0 in the broad energy range 1–50 keV (SXT: 1-5 keV and LAXPC20: 4-50
keV) for 4,  6 and  7 sets of observations listed in Table \ref{tab:table1}. The high energy range above 50.0 keV has been ignored because of the low S/N (signal-to-noise) ratio. For the rest of the observation sets, we have used the combined SXT, and LAXPC
energy range 1.0-20.0 keV; during these observations source spectrum is soft and signal to noise ratio deteriorates fast above 20 keV. Lower energies below 1 keV were not considered in all the observations due to uncertainties in the effective area and response of the  SXT. The left panels of Fig. \ref{fig:fig5} display the energy spectra of HS state (Obs 3, Seg. 2) in the top panel and SPL state (Obs. 1) in the bottom panel,  respectively, covering an energy range of 1-20 keV. The low HIMS and LH state spectra for the Obs. 4 (Seg. 6) and Obs. 7 (Seg. 6), respectively, are shown in the right top and right bottom panels of Fig. \ref{fig:fig5} in the energy range of 1-50 keV.  A relative normalisation constant was used for the simultaneous fitting of LAXPC and SXT data. 
As recommended by the LAXPC team, the  3\% systematic error was incorporated for uncertainties in background estimation when fitting LAXPC and SXT data together \citep[][]{antia2017calibration}.

\begin{figure*}
    \centering
    \includegraphics[scale=0.6]{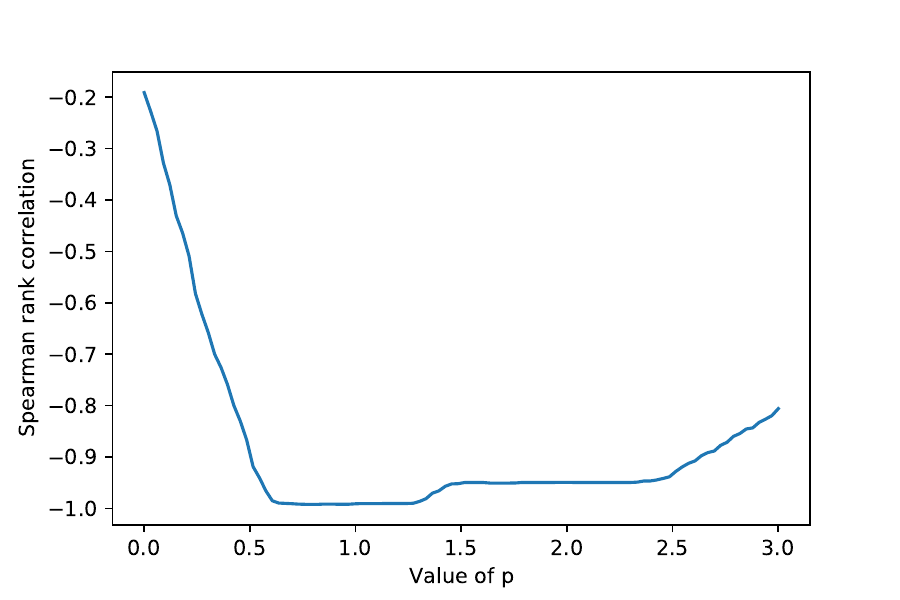}
    \caption{Variation in spearman rank correlation coefficient for $Y = \frac{\textit{QPO freq.}}{(\dot{M}^p)}$ with inner disc radius as a function of $p$. }
    \label{fig:fig7}
\end{figure*}

A gain correction was applied to the SXT data using the gain fit in XSPEC with slope fixed to 1, and the best-fit offset value was found to range from 0 to 35.68 eV.
SXT data were grouped with the ftgrouppha\footnote{\href{https://heasarc.gsfc.nasa.gov/lheasoft/ftools/headas/ftgrouppha.html}{https://heasarc.gsfc.nasa.gov/lheasoft/ftools/headas/ftgrouppha.html}} tool of Ftools\footnote{\href{https://heasarc.gsfc.nasa.gov/ftools/}{https://heasarc.gsfc.nasa.gov/ftools/}}. There are several
ways for binning the input pha file data; we have done the optimal binning
using the ftgrouppha tool. The spectrum was fitted using a combination of models,
Constant*tbabs (kerrdisk+simpl*kerrd). The absorption by the Inter-Stellar Medium (ISM)
was taken into account with the TBabs model \citep{wilms2000absorption}  implemented with the galactic absorption abundance. The hydrogen column density was kept fixed at $4 \times 10^{22}cm^{-2}$ for data sets of HIMS, SPL and LH states listed in Table \ref{tab:table3}, as there was no significant difference in the best-fit while keeping this parameter free \citep[][]{misra2020identification, liu2021testing}. $N_h$  was
kept free for HS state data set and was found to vary from $4.47 \times 10^{22}cm^{-2}$ to $4.65 \times 10^{22}cm^{-2}$. The convolution model of comptonization “simpl” \citep[][]{steiner2009simple} was used to take into account the Comptonization of the disk photons in the inner flow. The simpl model processes any input spectrum and transforms a fraction $f_{sc}$ of the source photons into a power-law distribution.
The inner radius of the disk and mass accretion rate was estimated from the best-fit values obtained from the relativistic disk model, “kerrd” \citep[][]{ebisawa2003accretion}. The black hole mass, disk inclination angle, and distance to the source were fixed to 12.4M$_{\odot}$, 60$^{\circ}$, and 8.6 kpc, respectively, \citep[][]{reid2014parallax}. The spectral hardening factor of kerrd was fixed to
1.7 \citep[][]{shimura1995spectral}. For the kerrdisk model, the emissivity index for both the inner and the outer portions of the disk was fixed at 1.8 \citep[][]{blum2009measuring}. The rest-frame energy of the iron line was set at 6.4 keV \citep[][]{blum2009measuring}. As GRS 1915+105 is a highly spinning

\begin{figure*}
    \centering
    \includegraphics[scale=0.6]{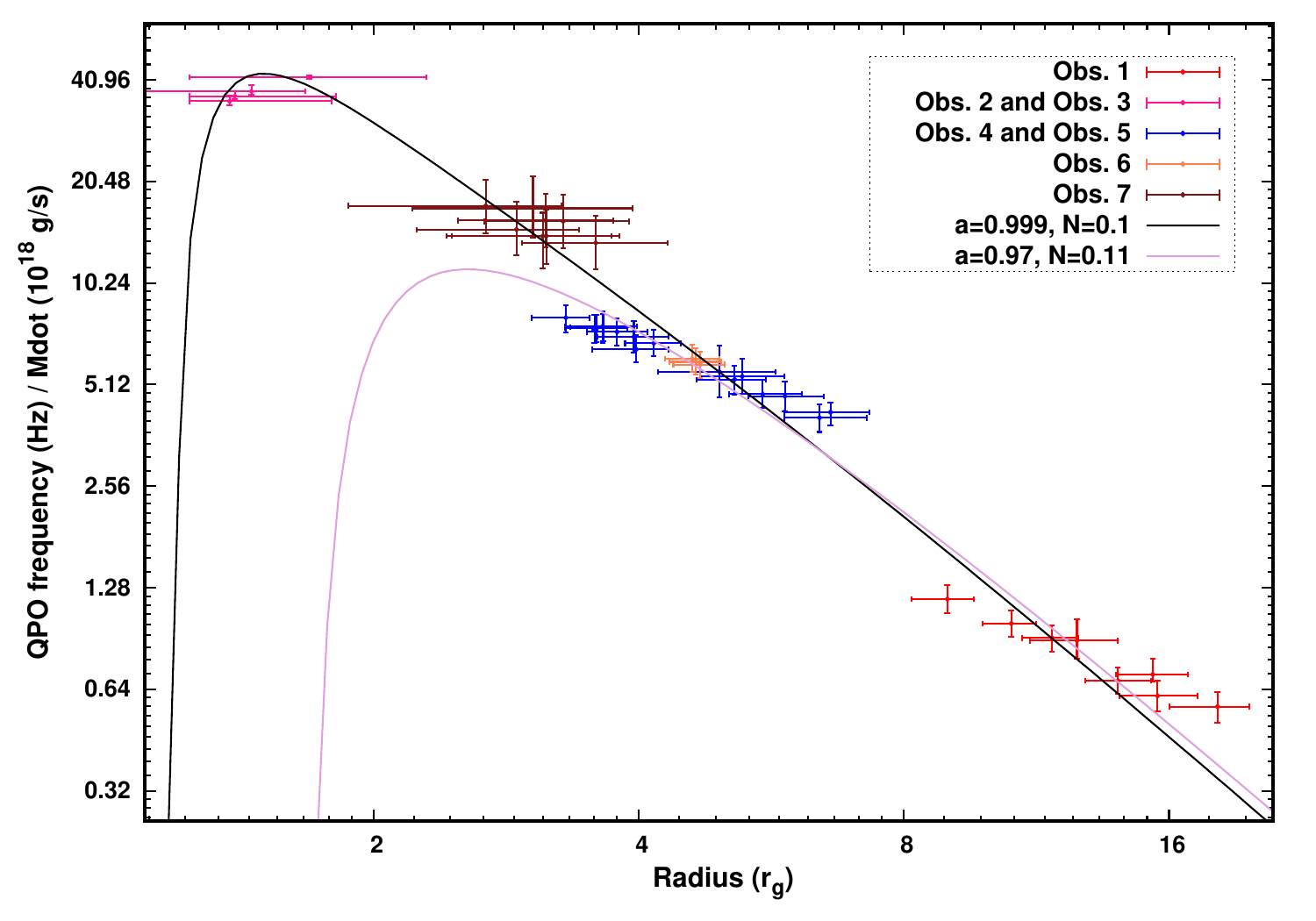}
    \caption{ QPO frequency divided by accretion rate with a broad vast of inner disk radius. The Black solid line indicates the relativistic standard accretion disc model for dimensionless spin parameter a = 0.999 and Normalisation constant N=0.1. However, the solid Plum colour line reflects the best-fit model presented by \citet{misra2020identification} for the Low-Frequency QPOs, where the best spin parameter was 0.973 and N value was 0.11.}
 \label{fig:fig8}    
\end{figure*}

black hole, we set the spin parameter for “kerrdisk” at 0.98 \citep[][]{blum2009measuring}. Keeping these parameters free does not significantly affect the best-fit values of other parameters. The break radius separating the inner and outer parts of the disk was fixed at 6 $r_g$ (gravitational radii). The radius parameter in kerrd is measured in the unit of gravitational radius ($r_g$), while for the kerrdisk, it is in units of radius of marginal stability or innermost stable circular orbit (ISCO). Therefore inner radius in the kerrdisk was normalised to that used for the “kerrd” after dividing by a factor of 1.235. The fraction scatter parameter in the data from 3 March 2016 was not constrained; therefore, we set it to 0.6. For HS state observation, gamma and flux in line emission parameters were not constrained; thus, we set them to $4.5$ and $ {1 \times 10^{-2}}\textrm{photons}$ $  \textrm{cm}^{-2}  \textrm{s}{-1}$, respectively. Table \ref{tab:table2} represents the best-fit values of the spectral parameters, including the absorption column density, inner disk radius, accretion rate, scattered fraction, photon index (gamma), and flux in the iron emission line.

\section{RESULTS}
\label{sec:results}

An overview of the observations used in this work which includes the date of observation, X-ray flux, hardness ratio, X-ray state, QPO frequency, accretion rate, and the inner disk radius, is given in Table \ref{tab:table3}. 
The X-ray flux observed in the LAXPC20 detector is presented in Column 2 of Table \ref{tab:table3}. The value of HR2 is shown in column 3, where HR2 is defined as the ratio of X-ray flux in the 13–60 keV to the 3-5 keV energy range. We observe that the hardness ratio continuously decreases as the source moves from the Low Hard (LH) state to the HS state via the SPL state and the low HIMS state. The accretion rate, shown in column 6 of Table \ref{tab:table3}, generally increases as energy spectra become softer. The accretion rate is highest during the  SPL state and lowest during the LH state.
Columns 5 and 7 of Table \ref{tab:table3} list the range of  QPO frequencies and the inner radii of the truncated disc for different observations.

Fig. \ref{fig:fig6} shows the variation of QPO frequency with accretion rate (top left panel), with inner disc radius (top right panel), and the variation of accretion rate with the inner disc radius (bottom panel). While for some of the individual data sets (i.e. for observations taken during a particular spectral state, such as Obs. 4 and Obs. 1), correlations between these parameters are evident, there is, in general, no correlation seen when all the observations are considered.

\begin{table*}
\centering
\caption{Overview of the analysis of GRS 1915+105 done in this work.}
\label{tab:table3}
\begin{tabular}{M{2.1cm} M{2.6cm} M{1.2cm} M{2.1cm} M{2.1cm} M{2.1cm} M{2.1cm} } 
\hline \hline
Observations  & X-ray flux for lxp20 (c/s) & HR2 $\left(\frac{13-60 keV}{3-5 keV}\right)$ & X-ray State & QPO freq.   (Hz) & Accretion rate ($ 10^{18}$ gm/s) & Inner radii $(R_g)$ \\
\hline

\vspace{4pt} Obs. 1 \vspace{4pt} & $\sim$ 1840 & $\sim 0.36$  & SPL / HIMS (high) & 2.56 - 4.94 & 4.17 - 5.25 & 8.96 - 18.15  \\ 
\hline 
\vspace{4pt} Obs. 2 \& Obs. 3 & $\sim$ 3324 &  $\sim$ 0.13 & HS & 69.81 - 71.43 & 1.69 - 1.97 & 1.37 - 1.69  \\
\hline 
\vspace{4pt} Obs. 4 &   $\sim$ 850 & $\sim$0.64  &HIMS (low)& 3.44 - 5.41 & 0.66 - 0.86  & 3.31 - 6.60 \\ 
\hline 
\vspace{4pt} Obs. 5 & $\sim$ 780 & $\sim$ 0.57  &HIMS (low)& 4.04 - 5.31 & 0.70 - 0.86  & 3.78 - 5.86 \\ 
\hline 
\vspace{4pt} Obs. 6 &   $\sim$ 720 & $\sim$ 0.46  &HIMS (low)& 4.48 - 4.52 & 0.74 - 0.77  & 4.60 - 4.70 \\ 
\hline 
\vspace{1pt} Obs.7 \vspace{4pt} & $\sim$ 320 & $\sim$ 0.99  & LH & 2.01-2.42 & 0.11-0.16  & 2.35 - 3.58 \\ 

\hline  

\end{tabular}
\end{table*}

Next, we consider the possibility that the QPO frequency may depend both on the accretion rate and the inner disc radius and, in particular, in the form suggested by Equation \ref{primeqn}, i.e. the QPO frequency divided by the accretion rate depends on the inner disc radius, as was suggested by \citet{misra2020identification}. This is illustrated in Fig. \ref{fig:fig8}, where the QPO frequency divided by the mass accretion rate is plotted against the inner radius of the accretion disc. In this case, a clear trend is visible for all the observations.  The solid violet line in Fig. \ref{fig:fig8} represents the best-fitted standard accretion disc model for Low Frequency QPOs (LFQPOs) with spin parameter 0.97 and normalisation constant 0.01 (earlier work; \citet{misra2020identification}, who used only low HIMS data). For all the data sets, we find that the relationship is consistent with that predicted by the dynamic frequency model (given in Equation 1 with $a=0.999$ and $N=0.1$). This is shown by the solid black in Fig. 8. Note that the high spin value is already implied by the small inner radii of $\sim 1.2 R_g$ obtained from the spectral fitting. This work extends the earlier results to different spectral states and covers a large variation in accretion rate from $0.1 \times 10^{18} \textrm{gm/s}$ to $5.0 \times 10^{18} \textrm{gm/s}$ and the truncated radius changing from the last stable orbit of a maximally spinning black hole, $\sim 1.2$ to $\sim 19$  Gravitational radii. For this wide range, the frequencies of the C-type QPO  follow the trend predicted by the relativistic model and, interestingly, the high frequency QPO at $\sim 70$ Hz (which is an obvious outlier in top panels of  Fig. \ref{fig:fig6}) also follow the same trend, suggesting a common origin.  While the qualitative trend is as predicted, there are quantitative deviations, which we discuss in the next section.

We have so far studied the QPO frequency divided by $\dot M$ as a function of the inner disc radius based on the interpretation that the QPO frequency is the dynamical one given by Equation \ref{primeqn}. To generalise, we define a variable $Y = \frac{\textit{QPO freq.}}{(\dot{M}^p)}$ and check if other values of $p$ other than unity would also represent the data by checking if $Y$ is correlated with inner disc radius. The absolute magnitude of the Spearman rank correlation has a maximum of 0.99 for $p$ ranging between 0.8 and 1.2. The Spearman rank correlation variation with $p$ is plotted in Fig. \ref{fig:fig7}. This figure shows that the correlation does not show significant change for $p$ values within 0.8 to 1.2.


\section{Discussion}
In order to put the results of this work into perspective, it is necessary first to enumerate the various possible different reasons why the data points in Fig. \ref{fig:fig8}, show some deviations from the predicted values. It has been assumed that the colour factor $f$ is a constant $=1.7$. The colour factor depends on the local vertical radiative transfer in the disc and has been numerically obtained to be approximately 1.7 by \citet{shimura1995spectral} for black hole binaries. The radiative transfer depends on the vertical structure of the disc and on the fairly uncertain viscous energy dissipation as a function of height. Moreover, a corona on top of the disc and irradiation will also affect the colour factor. The effect of changing the colour factor is more prominent for observations with a larger inner truncated disc radius. For example, if the colour factor is increased to 2, the mass accretion rates and the inner radii of the accretion disk slightly change for the soft state data collected on 25 April 2016 and 27 April 2016 i.e. mass accretion rate changes from $1.95 ^{+0.06}_{-0.02}$ to $ 1.93 ^{+0.10}_{-0.048} \times 10^{18}$ g/sec and the inner radius changes from $1.40 ^{+0.42}_{-0.15}$ to $1.32 ^{+0.62}_{-0.08} R_g$. On the other hand, for the Low HIMS (15 Apr 2017), the accretion rate change from $0.74 ^{+0.07}_{-0.06}$ to $2.4 ^{+0.3}_{-0.2} \times 10^{18}$ g/sec while the inner radius changes from $4.6^{+0.3}_{-0.3}$ to $9.6^{+1.0}_{-0.3} R_g$. An increase in the colour factor results in an increase in accretion rate and inner radii, making the HIMS points (Obs. 4, 5, 6) in Fig. \ref{fig:fig8} to move right and downwards. We have tested that by changing the colour factor to 2, then the predicted curve matches with the data points, but 
 the normalisation factor increases from 0.1 to 0.15. Note that we have also assumed that the colour factor is independent of the accretion rate and radii which may not be the case. Some of the deviations of the data points from the predicted values could be due to such dependence.

It should be emphasised that the theoretical formula for the dynamical frequency (Equation \ref{primeqn}) is an order of magnitude estimate, the uncertainty of which is parameterised by the normalisation factor $N$. Thus, one may expect $N$ to vary not only for different observations (with different accretion rates and inner disc radii) but also to vary with radius, leading to deviations when the data is compared with a constant $N$ prediction. The theoretical prediction is based on the standard accretion disc, where the disc extends to the last stable orbit and is not truncated. The sound speed at a radius may differ when the disc is truncated at that radius compared to when it is not, and this difference may be a function of the accretion rate and radius. A related issue is the assumption of standard accretion disc theory that the viscous dissipation goes to zero at the last stable orbit, which is incorporated both in the form of Equation \ref{primeqn} and in the spectral model {\it kerrbb} used in this work. This assumption forces the temperature (and hence the sound speed) to go to zero at the last stable orbit. However, this assumption may not correctly describe the system, and instead, the accretion flow should necessarily pass through a sonic point, which leads to deviations from the standard theory near the last stable orbit \citep[][]{abramowicz2013foundations}. Apart from these theoretical considerations, another potential reason for the deviation between the data and the predicted values is that the source may not be in the steady state and may be in a variable state. Out of seven observations used in this work,  the source 
shows significant short-time variability  {(on the hour/orbital time scale)}  during three observations  (3rd March 2016, 28th March 2017 and 1st April 2017  (Obs. 1, 4 \& 5)) \citep{yadav2016astrosat,rawat2018study}, as reflected in Table 2. During these observations, values of QPO frequency, inner disk radii and the Gamma clearly show a trend with time (for different orbits). Thus, the spectra averaged over the whole observation may not provide accurate accretion rates and inner disc radii values. Moreover, when the system is dynamic, it may not be correct to model the time-averaged spectra with a steady state one, as assumed when we have used a disc model like {\it kerrbb}. These three data sets show  most deviations from the theory, as seen in Figure \ref{fig:fig8} as the disk was not in the steady state. The  15th April 2017 (Obs. 6) data support this argument. This observation data do not show any trend with time/orbit and fall in the middle of points of Obs. 4 \& 5 in  Figure \ref{fig:fig8} with little deviation (also see Table \ref{tab:table2}). 

Given all the above-listed possibilities, which may cause the data points not to follow the theoretical predictions accurately, it is quite remarkable that the overall predicted trend is seen, for such a wide range of accretion rates, inner disc radii and QPO frequency. Indeed, as mentioned earlier, the general trend that for an empirical form of $Y = f_{QPO}/\dot M^p$, the best anti-correlation with radii is obtained for $p \sim 1$, indicates that the QPO frequency can be identified with dynamical one. It is also remarkable that the high frequency QPO at $\sim 70$ Hz also follows the trend of the low frequency ones and the explanation for the observed high frequency is that for the high frequency QPO, the accretion rate is significantly higher and the inner radius close to the last stable orbit.

Interpreting the QPO frequency as the dynamic one, is an alternate explanation to the model where the QPO is due to the precession of the inner flow at the Lense-Thirring frequency \citep[][]{stella1997lense, ingram2009low, ingram2019review, motta2022black, motta2016quasi, you2020x}. In that interpretation, the QPO frequency is expected to be a function only of the truncation radius and not the accretion rate. Moreover, there is some evidence that the energy dependent properties of some of the QPOs vary with the inclination angle of the binary \citep[][]{van2016inclination, motta2015geometrical, schnittman2006precessing, heil2015inclination, arur2020likely}, which would be more likely explained by a precessing inner flow. At present, this evidence is limited to a few sources due to the difficulty in estimating the inclination angle and energy dependent QPO properties. A more detailed theoretical analysis of the predicted inclination dependence of these two interpretations, along with better data, would be able to differentiate between them. Note that in the interpretation used in this work, the QPO frequency is not expected to depend on the inclination angle of the disc.
The wide band spectral and rapid temporal capabilities of AstroSat and Insight-HXMT had shown that the frequencies of the  C-type QPO of GRS 1915+105 can be identified with general relativistic dynamic ones. In this work, we extend the results using AstroSat for a broader range of accretion rates and inner radii and have shown that the high frequency QPO may also be of a similar origin. The work needs to be extended to other observations of GRS 1915+105 and other black hole systems. Apart from  AstroSat and  Insight-HXMT observations, such work can also be done by NICER with perhaps high energy spectral coverage from simultaneous Nustar data.  Such a systematic and multi-observatory study will give a clearer picture of the origin of the QPO phenomenon in black hole systems.

\section*{Acknowledgements}
\label{sec:ack}

The authors would like to thank the anonymous reviewer for his or her insightful remarks and suggestions that considerably enhanced the quality of the manuscript.
This work has used the data from the Soft X-ray Telescope (SXT) developed at TIFR Mumbai. And the SXT POC at TIFR is acknowledged for verifying and releasing the data through the Indian Space Science Data Centre (ISSDC) and providing the required software tools.
We would also like to thank the LAXPC POC and SXT POC teams for their support. In addition, this study utilised the Monitor of All-sky X-ray Image (MAXI) and SWIFT/BAT data provided by the MAXI and BAT teams.

This research has used the software provided by the High Energy Astrophysics Science Archive Research Center (HEASARC), a service of the Astrophysics Science Division at NASA.


\section*{Data Availability}
\label{sec:data_availability}

The software and packages utilised for data analysis are available at NASA’s HEASARC website (\url{https://heasarc.gsfc.nasa.gov/docs/software/heasoft/patch.html}). The data used \makebox[\linewidth][s]{in this article are available at the AstroSat-ISSDC website}  \\ (\url{https://astrobrowse.issdc.gov.in/astro_archive/archive/Home.jsp}), Maxi website (\url{http://maxi.riken.jp/top/index.html}) and the Swift/BAT observations from NASA’s SWIFT website 
(\url{https://swift.gsfc.nasa.gov/results/transients/}).


\bibliographystyle{mnras}
\bibliography{example}




\section{Appendix}

The relativistic correction parameters A, B, D, E, and L \citep[][]{novikov1973astrophysics, page1974disk} which have been used to derive equation 1 are as follows. 
\begin{align}
    &A=1+a_{*}^2x^{-4}+2a_{*}^2x^{-6}\nonumber\\&B=1+a_{*}x^{-3}\nonumber\\&D=1-2x^{-2}+a_{*}^2x^{-4}\nonumber\\&E=1+4a_{*}^2x^{-4}-4a_{*}^2x^{-6}+3a_{*}^4x^{-8}
\end{align}
\begin{align}
    L=\frac{3}{2M}&\frac{1}{x^2(x^3-3x+2a_{*})}\Biggl[x-x_0-\frac{3}{2}a_{8}\ln{\biggl(\frac{x}{x_0}\biggr)}\nonumber\\ &-\frac{3(x_1-a_{*})^2}{x_1(x_1-x_2)(x_1-x_3)}\ln{\biggl(\frac{x-x_1}{x_0-x_1}\biggr)}\nonumber\\ &-\frac{3(x_2-a_{*})^2}{x_2(x_2-x_1)(x_2-x_3)}\ln{\biggl(\frac{x-x_2}{x_0-x_2}\biggr)}\nonumber\\ &-\frac{3(x_3-a_{*})^2}{x_3(x_3-x_1)(x_3-x_2)}\ln{\biggl(\frac{x-x_3}{x_0-x_3}\biggr)}\Biggr]
\end{align}
where $x=\sqrt{\frac{r}{M}}$ and $a_{*}$ is the spin parameter in parameters A, B, D, E, L. Here,
\begin{align}
\hspace{2cm}&x_1=2\cos\Bigl({\frac{1}{3}\cos^{-1}{a_{*}}-\frac{\pi}{3}}\Bigr)\nonumber\\ &x_2=2\cos\Bigl({\frac{1}{3}\cos^{-1}{a_{*}}+\frac{\pi}{3}}\Bigr)\nonumber\\ &x_3=-2\cos\Bigl({\frac{1}{3}\cos^{-1}{a_{*}}}\Bigr)\nonumber\\ &x_0=\bigl\{3+Z_2-sgn(a_{*})[(3-Z_1)(3+Z_1+2Z_2)]^{\frac{1}{2}}\bigr\}^\frac{1}{2}
\end{align}
where $Z_1=1+(1-a_{*}^2)^{1/3}[(1+a_{*})^{1/3}+(1-a_{*})^{1/3}]$ and $Z_2=(3a_{*}^2+Z_1^2)^{1/2}$
\label{lastpage}
\end{document}